\documentclass[11pt]{JHEP3}
\usepackage{epsfig}
\usepackage{amsfonts}
\usepackage{latexsym}
\usepackage{amsmath}
\usepackage{datetime}

\usepackage{graphicx}
\usepackage{multirow,cite}
\usepackage{subcaption}

\newcommand{\bi}{\begin{itemize}}
\newcommand{\ei}{\end{itemize}}
\newcommand{\non}{\nonumber}
\def\p{\partial}
\def\a{\alpha}
\def\b{\beta}
\def\d{\delta}
\def\g{\gamma}

\def\e{\epsilon}

\def\om{\omega}

\def\A{\mathcal{A}}
\def\O{\mathcal{O}}

\def\G{\Gamma}
\def\V{\mathcal{V}}

\def\rp2{\mathbb{R}\mathrm{P}^2}

\def\r{\rightarrow}
\def\half{{\frac12}}

\newcommand{\bea}{\begin{eqnarray}}
\newcommand{\eea}{\end{eqnarray}}
\newcommand{\be}{\begin{equation}}
\newcommand{\ee}{\end{equation}}

\title{\vspace{-3cm} Behind the geon horizon}
\author{Monica Guica$^{\flat,\dag}$ and Simon F. Ross$^\sharp$\\

\vspace{1mm}
\hspace{-4.5mm}{\small $ {}^\flat$   Nordita, KTH Royal Institute of Technology and Stockholm University,\\
\hspace{-0.35 cm}
Roslagstullsbacken 23, SE-106 91 Stockholm, Sweden\\
}

\hspace{-4.5mm}{\small $ {}^\dag$   Department of Physics and Astronomy, Uppsala University,\\  \hspace{-0.35 cm}  SE-751 08 Uppsala, Sweden\\
}

\hspace{-4.5mm}{\small $ {}^ \sharp$  Centre for Particle Theory, Department of Mathematical Sciences,
\\ \hspace{-0.35 cm} Durham University, South Road, Durham DH1 3LE, UK\\
}}
\abstract{
\vspace{0.5 cm}

 We explore the Papadodimas-Raju prescription for  reconstructing the region behind the horizon of one-sided black holes in AdS/CFT in the case of the $\rp2$ geon - a  simple, analytic example of a single-sided, asymptotically AdS$_3$ black hole, which corresponds to a pure CFT state that thermalizes at late times. We show that in this specific example, the mirror operators involved in the reconstruction of the interior have a particularly simple form: the mirror of a single trace operator at late times is just the corresponding single trace operator at early times. We use some explicit examples to explore how changes in the state modify the geometry inside the horizon.  
 }

\begin{document}

\newcommand{\todo}[1]{{\em \small {#1}}\marginpar{$\Longleftarrow$}}   

\section{Introduction and summary}

There has recently been considerable renewed interest in the issues of principle raised by black hole evaporation and the information loss problem, initiated by the firewall argument \cite{Almheiri:2012rt,Almheiri:2013hfa} (see also \cite{Braunstein:2009my,Mathur:2009hf}), which uses insights from quantum information theory to sharpen the tension between the existence of a smooth horizon and unitarity of the evaporation process (for a useful review, see  \cite{Harlow:2014yka}). An important contribution to these developments was the proposal of Papadodimas and Raju (henceforth PR) \cite{Papadodimas:2012aq,Papadodimas:2013wnh,Papadodimas:2013jku} of a  concrete recipe  for the reconstruction of the region behind the horizon of one-sided black holes - dual to some pure state $| \Psi \rangle$ - from the point of view of the dual CFT. Their prescription is based on identifying a set of operators in the CFT - called ``mirror operators'' - which are entangled with the degrees of freedom  corresponding to fields outside the horizon at late times in a way analogous to the entanglement between the two boundaries of an eternal black hole \cite{Maldacena:2001kr}.\footnote{See also \cite{Verlinde:2012cy} and the ER=EPR  proposal \cite{Maldacena:2013xja} of a  connection between entanglement and the geometry of the eternal black hole.} These operators commute with operators describing the black hole exterior inside all low-point correlation functions, but not exactly - an essential feature that allows the implementation of black hole complementarity consistently with local effective field theory in the bulk. In the PR proposal, local bulk fields are reconstructed from generalized free field operators in the boundary CFT and their mirrors
in the same way as the interior of the eternal black hole can be described in terms of the operators on the two boundaries \cite{Banks:1998dd,Balasubramanian:1999ri,Bena:1999jv,Hamilton:2005ju,Hamilton:2006az,Hamilton:2006fh}. 

The PR construction assumes that there is a smooth horizon, and that the description of the region behind the horizon is essentially the same as in the eternal black hole. A key question is,  for which states in the dual CFT is such an assumption appropriate. PR argue that the prescription applies to ``equilibrium" states, i.e. states that look thermal from the point of view of  a small subset of observables $\mathcal A$ used to probe the system. In our case, the condition is that correlators of operators in $\A$ should be indistingushable from thermal correlators, up to corrections exponentially suppressed in the entropy of the thermal state:
\begin{equation} \label{tcond}
\langle \Psi |A_p | \Psi \rangle = \mbox{Tr} (\rho_{th} A_p) + \O(e^{-c \, S_{\rho}})\;, \quad \quad \forall A_p \in \mathcal A
\end{equation}
where $\rho_{th}$ is a thermal density matrix and $c$ is an order one constant.\footnote{More generally, one would have a similar condition with an appropriate ensemble density matrix replacing $\rho_{th}$, for example the microcanonical ensemble.}  Given an equilibrium state, PR define the mirror operator $\tilde \O_\Psi$ of an operator $\O \in \mathcal A$ to be an operator satisfying 
\begin{equation} \label{conj}
\tilde{\mathcal O}_\Psi |\Psi \rangle = e^{-\beta H/2} \mathcal O^\dagger e^{\beta H/2} |\Psi \rangle, 
\end{equation}
\begin{equation} \label{comm}
[ \tilde{\mathcal O}_\Psi,A_p] |\Psi \rangle = 0, 
\end{equation}
where $A_p$ is any  operator in $\mathcal A$ and the subscript on $\tilde \O_\Psi$ emphasizes the fact that the mirror operator depends on the reference state $|\Psi \rangle$. PR showed that a solution to these equations always exists, provided that no combination of operators in $\A$ annihilates $| \Psi \rangle$, which is usually the case  if the algebra $ \mathcal A $ is  small and the state $| \Psi \rangle$ is sufficiently complicated.

States in the CFT generated by acting on an equilibrium state $|\Psi \rangle$ with a CFT operator are supposed to correspond to some bulk excitation on top of the geometry dual to $|\Psi \rangle$. To the extent that such out-of-equilibrium states can be detected by the observables in $\A$, the PR construction can be easily modified to account for these situations \cite{Papadodimas:2013jku}. 

The PR proposal has been the subject of some controversy: \cite{Almheiri:2013hfa, Harlow:2014yoa} constructed examples of states satisfying the equilibrium condition which can also be realized as excitations of another equilibrium state, thus questioning the well-posedness of the  construction. Important issues of principle are also raised by the state-dependence of the mirror operators \cite{Almheiri:2013hfa, Harlow:2014yoa}. In addition, the definition of the mirror operators is indirect, so it is difficult to gain intuition into their nature  in a generic pure state. It may be enlightening to explore these issues in the context of an explicit example, where the mirror operators have a simple interpretation.

In this paper we study the simplest example of a one-sided black hole spacetime, the $\rp2$ geon, which is obtained by a quotient of the BTZ black hole \cite{Louko:1998hc}. An advantage of this example is that the dual CFT state is explicitly known \cite{Maldacena:2001kr}. The bulk geometry has a smooth horizon and, as we will see, if we take $\mathcal A$ to consist of correlation functions of 
local operators at late times, then the equilibrium condition \eqref{tcond} is satisfied.

Our first aim is to explicitly identify  PR's mirror operators for this case. We  argue that, given a late-time local operator $\mathcal O(t,\phi)$, the corresponding local operator $\mathcal O(-t,\phi+\pi)$ acting at early times satisfies the requirements \eqref{conj} - \eqref{comm} for a mirror operator, so long as $t > t_*$, the scrambling time for the bulk black hole. Thus, the mirror operators can be identified, at least at leading order in the central charge, with simple local operators in the CFT. We note that as in the eternal black hole, it is natural to think of the state as specified on the $t=0$ surface in the bulk, so these early-time operators can be thought of as precursor operators in the sense of  \cite{Susskind:2013lpa,Susskind:2013aaa}.

We  also study some simple examples of changes to the geon state generated by unitary transformations that do not affect the equilibrium condition \eqref{tcond}. If we assume that the resulting states are dual to geometries with a smooth horizon, there is a corresponding change in the boundary conditions inside the horizon in the bulk. This illustrates how the change in state can be interpreted as a modification of the geometry behind the horizon.  

This paper is organised as follows. In section \ref{review}, we review the geometry of the $\rp2$ geon space-time as a quotient of BTZ and argue, using  holographic computations, that the dual state satisfies \eqref{tcond} with respect to late-time correlation functions. In section \ref{pathintg}, we  review the path integral construction of the state dual to the geon following \cite{Maldacena:2001kr}, and use this to discuss the entanglement structure 
and high-energy support 
of the geon state from a CFT perspective. In section \ref{mirror}, we review the construction of local bulk operators in BTZ and introduce a smeared version thereof. We then extend this construction to  bulk fields in the geon spacetime, and use it to support our identification of the mirror operators. Finally, in section \ref{deviations}, we discuss some examples of unitary rotations of the geon state. Several technical details are collected in the appendices.

\section{The $\rp2$ geon geometry}
\label{review}

In three dimensions, the absence of local gravitational degrees of freedom implies that the metric for a vacuum solution is locally AdS$_3$. A number of physically interesting examples can be constructed as quotients of global AdS$_3$, including the BTZ black hole \cite{Banados:1992wn,Banados:1992gq}, and a rich family of single-sided black holes \cite{Brill:1995jv,Aminneborg:1997pz,Brill:1998pr}, of which we will focus on the $\rp2$ geon as the simplest example \cite{Louko:1998hc}.

\subsection{Definition \label{defgeon}}

It is convenient to describe AdS$_3$ in embedding coordinates, as a hyperboloid $- T_1^2 - T_2^2 + X_1^2 + X_2^2 = - \ell^2$ in $\mathbb R^{2,2}$, with metric 
\begin{equation}
ds^2 = -dT_1^2 - dT_2^2 + dX_1^2 + dX_2^2.
\end{equation}
In these coordinates, the $SO(2,2) \simeq SL(2,\mathbb{R}) \times SL(2,\mathbb{R})$ isometries of AdS$_3$ are realised as Poincar\'e boosts in the $T_i, X_i$ coordinates. Global AdS$_3$ is the universal covering space obtained by unwrapping the angular coordinate in the $T_1, T_2$ plane to take all real values. The non-rotating BTZ black hole is obtained by restricting to the region $T_1^2 > X_1^2$, where the Killing vector
\begin{equation}
\xi = X_1 \partial_{T_1} + T_1 \partial_{X_1} \label{defxi}
\end{equation}
is spacelike, and quotienting by the discrete isometry group $\Gamma_\gamma \simeq \mathbb{Z}$ generated by $\gamma = e^{2\pi r_+ \xi/\ell}$. This quotient preserves a $U(1) \times U(1)$ subgroup of the 
original isometries, generated by $\xi$ and 
\begin{equation}
\chi = X_2 \partial_{T_2} + T_2 \partial_{X_2}.
\end{equation}
These isometries are manifest in  adapted coordinates, defined by 
\begin{equation} \label{t1x1}
T_1 = \frac{r \ell}{r_+} \cosh (\frac{r_+ \phi}{\ell}) \;, \quad \;\; X_1 = \frac{r \ell}{r_+} \sinh (\frac{r_+ \phi}{\ell}), 
\end{equation}
\begin{equation}
T_2 = \ell \sqrt{ \frac{r^2}{r_+^2} -1 } \sinh (\frac{r_+ t}{\ell^2})\;, \quad \quad X_2 = \ell \sqrt{ \frac{r^2}{r_+^2} -1} \cosh (\frac{r_+ t}{\ell^2}),\label{t2x2} 
\end{equation}
in terms of which $\xi = \frac{\ell}{r_+} \partial_\phi$, $\chi = \frac{\ell^2}{r_+} \partial_t$.  In these coordinates, the metric reads 
\begin{equation}
ds^2 = - \frac{r^2-r_+^2}{\ell^2} dt^2 + \frac{\ell^2}{r^2-r_+^2} dr^2 + r^2 d\phi^2 \label{btzmet}
\end{equation}
and the quotient generated by $\gamma$ acts as $\phi \sim \phi+ 2\pi$. This is the BTZ black hole \cite{Banados:1992wn,Banados:1992gq}, of inverse temperature $\beta = 2\pi \ell^2/r_+$ and entropy $S_{BH} = \pi r_+/2 G$. The quotient generates the full, maximally extended BTZ spacetime depicted in figure \ref{btzfig} below, which has two asymptotically AdS$_3$ regions. The adapted coordinates above cover just the exterior region on one side (region I) of the black hole. There are related coordinates which cover each of the other patches\footnote{In regions II, III, IV, one again defines coordinates $(t,r,\phi)$ that cover each of the respective patches only. The relation between $T_1,X_1$ and $r,\phi$ is the same as \eqref{t1x1}, only the relation between $T_2,X_2$ and $t,r$ changes: for region II, we need to exchange the formulae for $T_2 \leftrightarrow X_2$ in \eqref{t2x2}, in region III, we simply change the sign $T_2 \r - T_2$, $X_2 \r -X_2$, and  in region IV we both change the sign and interchange $T_2,X_2$. The metric always takes the form \eqref{btzmet}.
The sign change  $T_2 \r -T_2$, $X_2 \r - X_2$ can be implemented by the analytic continuation $t \r t + i \b/2$. Note that the Penrose diagram effectively captures the conformal dynamics in the $T_2,X_2$ plane. }; note that the time coordinate $t$ runs in opposite directions in the two asymptotic regions.

\begin{figure}[h]
\centering
\includegraphics[height=3.4 cm]{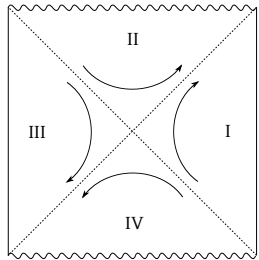} 
\caption{The Penrose diagram for the eternal BTZ spacetime, indicating the direction of $\chi = \partial_t$ in each region.}
\label{btzfig}
\end{figure}

\begin{figure}[h]
\centering
\includegraphics[height=4 cm]{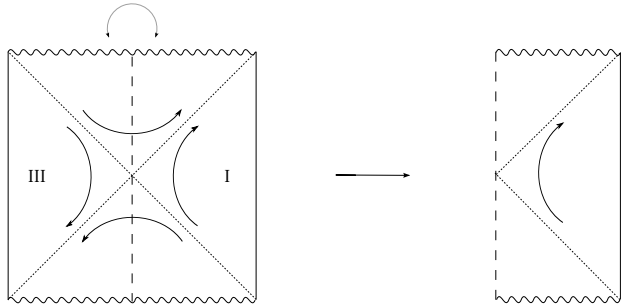} 
\caption{The construction of the geon spacetime as a quotient of BTZ. There is a time-translation symmetry that acts only in  the exterior region, which is  indicated by the arrow. }
\label{geonfig}
\end{figure}

It is easy to extend this analysis to generate geometries with a global event horizon and a single asymptotically AdS$_3$  region, by considering more complicated quotient groups \cite{Aminneborg:1997pz}. Here we will consider the simplest example, the $\rp2$ geon \cite{Louko:1998hc}, which is constructed by taking a further quotient by the $\mathbb Z_2$ action generated by 
$
\tilde J  = J \circ  e^{\pi r_+ \xi/\ell}
$,
where $J$ is the simple reflection
\be
J: X_2 \to -X_2. \label{geonq}
\ee
Since $\tilde J^2 = \gamma$, this generates a $\mathbb Z_2$ action on the BTZ spacetime, where we have already quotiented by $\G_\g$.  
In the BTZ coordinates of \eqref{btzmet}, $\tilde J$ maps a point at coordinates $(t, \phi)$ in region I to one at $(-t, \phi+\pi)$ in region III, and acts within regions II, IV to identify $(t, \phi) \sim (-t, \phi +\pi)$. Pictorially, this identifies points in figure \ref{btzfig} by a reflection around the vertical axis, accompanied by a rotation by $\pi$ in the circle direction, which is suppressed in the picture. The Penrose diagram of the geon is thus as shown in figure \ref{geonfig}. The $\phi$ rotation makes the spacetime smooth on the dashed axis on the left of this picture; the quotient by $\tilde J$ has no fixed points in the spacetime. 

The resulting spacetime is not orientable; the spatial slices have topology $\rp2$ minus a point, whence the name. However, in contexts where  AdS$_3$ is obtained by dimensional reduction from some higher-dimensional theory, one can construct an orientable spacetime in the higher-dimensional theory by combining \eqref{geonq} with an orientation-reversing involution of the internal space \cite{Louko:1998hc}.

In \cite{Maldacena:2001kr}, the quotient of BTZ by $J$ was also considered, which identifies $X_2 \r - X_2$, but without the additional shift in $\phi$. Unlike the geon, this quotient does have a fixed point at $X_2 =0$ ($t=0$ in regions II, IV) and the resulting space-time - which we will imaginatively denote as the ``$J$ quotient of BTZ''- has an orbifold singularity along this line. Nevertheless, it still represents a perfectly well-behaved, analytic single-sided black hole, whose Penrose diagram is again given by the left  figure \ref{geonfig}, the dashed vertical line now representing  the singularity. Region I of this spacetime, defined for $r > r_+$, is indistinguishable from its geon counterpart.

\subsection{Thermality of the  $\rp2$ geon }
\label{therm}

Since the geon spacetime only has one boundary,  the dual CFT state is expected to be pure;   the presence of a smooth horizon in the bulk indicates this  state should  thermalise. This picture is corroborated by the fact that  the geon space-time is not stationary - it does not admit a globally defined timelike Killing vector - and thus the dual state cannot be thermal at all times.  In this subsection, we  show that the geon state becomes thermal at late times, when probed by  correlation functions of CFT operators dual to local bulk fields. 
  
Since the exterior geometry of the geon is precisely the same as that of BTZ, the holographic calculation of one-point functions of boundary operators (at leading order in the central charge) will give exactly the same answer as in BTZ. Thus,  they will be thermal at all times. The departure from thermality is however reflected in the two- and higher-point functions of boundary operators, as the boundary-to-boundary propagator will be modified by the change in the bulk geometry inside the horizon. 

\begin{figure}[h]
\centering
\includegraphics[height= 3.4cm]{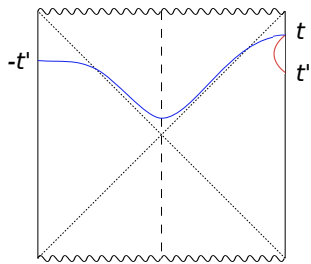}
\caption{The geodesics contributing to the two-point function of two operators on the same boundary (red), and for operators on different boundaries (blue) in BTZ. }
\label{btzgeod}
\end{figure}

Let us first consider the two-point function of free (scalar) fields propagating in the geon geometry, corresponding via the  AdS/CFT dictionary to generalized free field operators in the dual large $N$ CFT, whose correlation functions factorise \cite{ElShowk:2011ag}.   Since the geon space-time  is obtained from BTZ  by an involution, the bulk two-point function of free fields in the geon can be obtained from the one in BTZ by summing over the two images in BTZ of a given point in the geon,
\begin{equation} \label{twopt}
\langle \Phi( x) \Phi( x') \rangle_{geon} = \langle \Phi( x) \Phi( x') \rangle_{BTZ} +  \langle \Phi( x) \Phi(\tilde J ( x')) \rangle_{BTZ}. 
\end{equation}
For points that lie on the geon asymptotic boundary, the first term corresponds to the correlator of two operators inserted on the same BTZ boundary, while the second term receives contributions from  operators inserted on the two opposite boundaries of BTZ. The analyticity of the geon and BTZ spacetimes allows us to apply a geodesic approximation to calculate these boundary to boundary propagators for sufficiently heavy fields; the main contributions to the correlators on the RHS of \eqref{twopt} are well approximated by the geodesic length in the bulk, as depicted in figure \ref{btzgeod}. Explicit expressions for both correlators can be found in \cite{Louko:2000tp}.  The first contribution will be the same as the thermal BTZ two-point function, while the second contribution is sensitive to the non-thermal nature of the state. Near $t=0$, the second contribution in \eqref{twopt}, although smaller than the first, is not suppressed relative to it by any factor of the central charge, just by the difference in geodesic lengths \cite{Louko:2000tp}, and thus the resulting total  two-point function  is not  thermal.

\bigskip

 If we consider the equal-time two-point function at increasing $t$, the first contribution $\langle \Phi( x) \Phi( x') \rangle_{BTZ} $ is time-independent by the time-translation invariance of BTZ, but the contribution from $\langle \Phi( x) \Phi(\tilde J ( x')) \rangle_{BTZ}$ decreases with time. The relevant bulk geodesic for evaluating $\langle \Phi(x) \Phi(\tilde J ( x')) \rangle_{BTZ}$ is the one considered in the entanglement entropy calculation of \cite{Hartman:2013qma}, where is was found that the length of this geodesic increases linearly in time as $L \sim t/\beta$, so the contribution to the two-point function decreases as 
\begin{equation} \label{gp}
\langle \Phi( x) \Phi(\tilde J  x') \rangle_{BTZ} \sim e^{-\Delta' t/\beta }, 
\end{equation}
where $\Delta'$ is proportional to the mass of the bulk field in AdS units. Thus, at times large compared to the scrambling time \cite{Hayden:2007cs,Sekino:2008he}
\begin{equation}
t_* =\frac{ \beta}{2\pi} \ln S_{BH}
\end{equation}
the contribution from the geodesic that passes inside the horizon is exponentially suppressed in the black hole entropy $S_{BH}$. Its contribution is thus  smaller than than that from subleading bulk saddles we have neglected in our analysis, so at a time of this order we can say that the holographic two-point function has thermalised. 

We can easily extend the argument to perturbatively include interactions in the bulk. As in \cite{Papadodimas:2013jku}, we only consider  $n$-point correlation functions of operators dual to supergravity fields,  with $n$  finite. Thus,  the bulk backreaction can be neglected and the calculation is well approximated by quantum field theory on the geon geometry as a fixed background. Perturbative correlation functions in the bulk can be calculated by analytic continuation from the Euclidean geon spacetime. As in \cite{Kraus:2002iv}, this yields an expression for the Lorentzian correlation functions, in  which the interaction vertices are integrated only over the region outside the horizon, $r > r_+$. 

\bigskip

\begin{figure}[h]
\centering
\begin{subfigure}{5cm} 
\centering
\includegraphics[width=1.7cm]{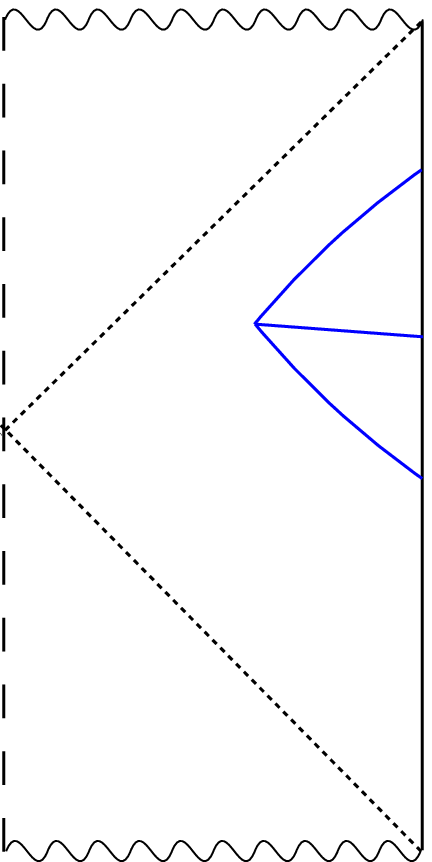}
\caption{}\label{corr-a}
\end{subfigure}
\quad \quad
\begin{subfigure}{5cm}  
\centering
\includegraphics[width=1.7cm]{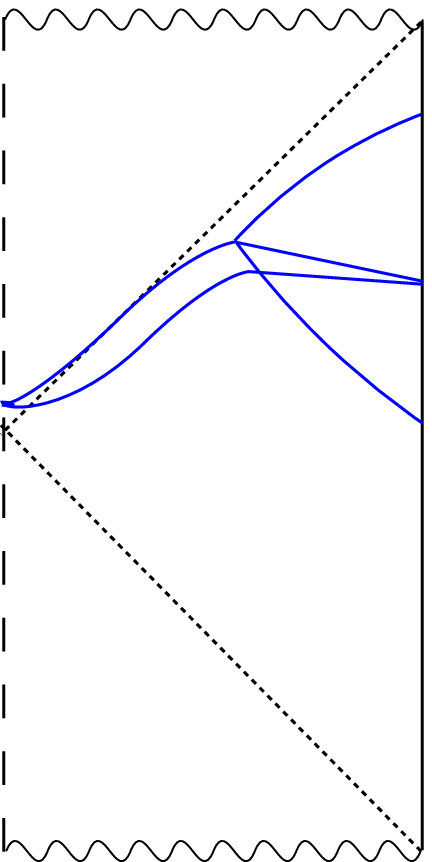}
\caption{}\label{corr-b}
\end{subfigure} 
\caption{Contributing bulk diagrams to the boundary three-point function. The interaction vertex is to be integrated only in the region outside the horizon.  }  
\end{figure}

The propagators connecting these vertices receive  contributions both from paths that do not pass through the identification, and  from paths that do, as in \eqref{twopt}. The perturbative contributions to the correlation functions can then be decomposed into contributions as in figure \ref{corr-a}, where no propagator passes through the identification, and those where one or more propagators does pass through the identification, as in figure \ref{corr-b}. The former diagrams give a thermal result, while the latter give a non-thermal contribution, so we would like to show that they are suppressed at late times. If we take the external points at late times on the boundary, the integration over the location of the interaction vertices has support mainly at late times, because the propagator in the exterior region $\langle \Phi(t) \Phi(t') \rangle$ is exponentially suppressed at large $t-t'$ due to dissipation. But for vertices at late times, the propagator through the identification $\langle \Phi( x) \Phi(\tilde J ( x')) \rangle$ is exponentially small by \eqref{gp}. Thus the contribution from diagrams as in  figure \ref{corr-b} is small at late times, and the correlation functions will be dominated by the thermal contributions from the diagrams in figure \ref{corr-a}. 

Thus, if we consider the set of observables $\mathcal A$ consisting of correlation functions of operators $\mathcal O$ dual to bulk supergravity fields (up to some finite order) at times $t > t_*$, the holographic calculation indicates that the state dual to the geon satisfies the equilibrium condition \eqref{tcond}. It is therefore a suitable candidate for applying the PR prescription.

\section{Path integral construction of the geon state \label{pathintg}}

The discussion in the previous section was entirely from the holographic point of view; however,  one of the advantages of the geon is that there exists  an explicit CFT path integral construction of the dual state, which we  review in this section. We also discuss the implications of this construction  for the  
 structure of the geon state.

In \cite{Maldacena:2001kr}, the CFT state dual to the BTZ black hole was identified using a Euclidean path integral construction. The initial data defining the black hole on a time slice $t=0$ can be obtained by a path integral over half of the Euclidean BTZ black hole, which yields the corresponding state. The Euclidean black hole is a solid torus, so half the boundary is a cylinder: the product of  a circle of length $2\pi$ parametrised by $\phi$  and of an interval of length $\beta/2$ parametrised by the Euclidean time, $\tau$. The dual state is  the so-called thermofield double state
\begin{equation} \label{tfd} 
|\Psi\rangle_{TFD} = \sum_i e^{-\beta E_i/2} |i\rangle_1 |i \rangle_2,
\end{equation}
which  belongs to the double copy of the CFT Hilbert space. In the above,  $| i \rangle_{1,2}$ are energy eigenstates of energy $E_i$ in the first/second copy of the CFT.

This construction can be generalised to any BTZ quotient  with a surface of time-reversal symmetry, since any such quotient will have a corresponding Euclidean spacetime continuation. The resulting state can again be identified by a path integral over half the boundary of the Euclidean space \cite{Krasnov:2000zq,Krasnov:2003ye,Skenderis:2009ju}; for the  $\rp2$ geon, it was obtained already in \cite{Maldacena:2001kr}. The quotient by $\tilde J$ acts on the Euclidean black hole by identifying $(\tau, \phi) \sim (\beta/2 - \tau, \phi + \pi)$.\footnote{This can be seen either directly from writing $\tilde J$ in the analytically continued BTZ coordinates, or by noting that in the Lorentzian spacetime it identifies $(t,\phi)$ in region I with $(-t,\phi+\pi)$ in region III, and that the BTZ coordinates in region III are related to those in region I by the analytic continuation $t \to t + i \beta/2$.} Thus, the boundary of the Euclidean continuation of the geon is a Klein bottle. Half the boundary is a M\"obius strip: starting from the cylinder of length $\b/2$ in the eternal black hole case, the two halves of the cylinder are identified, with the surface at $\tau = \beta/4$  identified with itself under a $\pi$ rotation. The CFT state dual to the $t=0$ initial data surface in the geon is thus given by the Euclidean path integral over the M\"obius strip. This gives for the geon state
\begin{equation} \label{geonstate}
|\Psi_g\rangle = e^{-\beta H/4} |C \rangle,
\end{equation}
where $|C\rangle$ is the cross-cap state  defined by the identification under $\phi \sim \phi + \pi$ of the CFT fields at $\tau=0$. 

\subsection{Entanglement structure \label{entstr}}

We would first like to use this expression for the geon state to understand its entanglement structure, which will be essential for the  identification of the mirror operators in section \ref{mid}. The geon state \eqref{geonstate} exhibits entanglement between different degrees of freedom in a \emph{single} copy of the CFT, in contrast to the entanglement between two copies of the CFT present in the thermofield double state \eqref{tfd}. This is the structure one would expect generically for states dual to a single-exterior black hole.  

The cross-cap state $|C \rangle$ is an entangled state between left- and right-moving modes. If we consider for example the free boson CFT, this state can be constructed explicitly. In terms of the  modes $j_n$ of the holomorphically conserved current $J = i \p X$  and $\bar j_n$ of the anti-holomorphic current $\bar J = i \bar \p X$  , the crosscap state can be shown to satisy (see e.g. \cite{Blumenhagen:2009zz})
\be
(j_n+(-1)^n \bar{j}_{-n})| C \rangle =0. \label{freecc}
\ee
The solution is
\be \label{freeexp}
|C \rangle = \exp \left(-\sum_{k=1}^\infty \frac{(-1)^k}{k} j_{-k} \, \bar{j}_{-k}\right) | 0 \rangle,
\ee
which shows perfect entanglement between the left- and the right-movers in the CFT. Alternatively, one can write this state as
\be
| C \rangle = \sum_{\vec{m}} |\vec{m} \rangle \otimes |  \overline{U_c \vec{m}} \rangle, \label{exprfbcc}
\ee
where 
\be
| \vec{m} \rangle = |m_1, m_2, \ldots \rangle = \prod_{k=1}^\infty  \frac{1}{m_k!} \left( \frac{j_{-k}}{\sqrt{k}}\right)^{m_k} |0\rangle,
\label{defvecm}
\ee
and $U_c$ is an anti-unitary operator, whose action on the current modes is  $U_c\,  j_n \, U_c^{-1} = - (-1)^n  j_n$.  
In more general CFTs, the cross-cap state satisfies 
\be
(L_n-(-1)^n \bar{L}_{-n})| C \rangle =0, \label{vircc}
\ee
where $L_n, \bar L_n$ are the Virasoro generators, so it again involves entanglement between left and right-movers. In particular, $ L_0 | C \rangle = \bar L_0 | C \rangle$.  However, the Virasoro algebra is not spectrum generating, so this condition cannot be directly solved to simply write the state as in \eqref{freeexp}. An expression for the crosscap state in terms of the so-called Ishibashi states is explicitly known in rational CFTs (see e.g. \cite{Blumenhagen:2009zz}).

In the free boson CFT, it is possible to show, using \eqref{freecc}, that simple operators  constructed from $J, \bar{J}$ and the vertex operators $\mathcal{V}_\a$ satisfy
\be
\phi_{h,\bar h}^\dag(t,\phi) |C \rangle=\bar{\phi}_{\bar h, h} (-t,\phi+\pi) | C \rangle,  \label{cccond}
\ee
where $h,\bar h$ represent the left/right conformal weights of the operator $\phi_{h,\bar h} $ and $\bar{\phi}_{\bar h, h}$ is an operator 
of the same dimension but of opposite spin to  $\phi_{h,\bar h}$,  obtained (in this particular case)  by the replacement $j_n \leftrightarrow \bar j_n$. 
 Given the geometrical construction of the crosscap state, it is natural to  assume that the above property generalizes to arbitrary operators in general CFTs\footnote{ In appendix  \ref{argcc}, we give an argument for its correctness in rational CFTs.}; all CFTs of interest will contain an operator $\bar \phi_{\bar h , h}$.

It is then easy to derive a similar relation for the geon state  \eqref{geonstate}. Letting $\phi(t,\phi) = e^{\b H/4} \O(t,\phi) e^{-\b H/4}$ in  \eqref{cccond}, where $\O(t,\phi)$ is some local operator in the CFT, we find that on the geon state
\be
e^{-\b H/2} \, \O^\dag(t,\phi)\, e^{\b H/2} | \Psi_g \rangle = \bar \O(-t,\phi+\pi) | \Psi_g \rangle \label{condo}
\ee
Note that the above relation takes the same form as  PR's conjugacy condition \eqref{conj}, with  $\bar{\O}(-t,\phi+\pi)$ playing the role  of the mirror operator\footnote{Note that in order for $\bar {\O} (-t,\phi+\pi)$ to be the mirror operator to $\O(t,\phi)$, it also needs to satisfy the commutation condition \eqref{comm}. We discuss when this is fulfilled in section \ref{mid}}. Interestingly, when the CFT$_2$  has an AdS$_3$ gravity dual and  $\O (t,\phi)$ is a generalized free field dual to a free scalar in the bulk, then the condition \eqref{condo} is precisely what one obtains for the Hartle-Hawking like state \cite{Louko:1998dj} upon quantizing the bulk field, as we review in section \ref{mid}. Our arguments above indicate that this relation continues to hold at all orders in $1/N$, where $N$ is the parametrically large central charge of the CFT.

If $\O(t,\phi)$ is a  generalised free field operator in a large $N$ CFT, then its Fourier modes $\O_{\om,m}$  are expected to obey a harmonic oscillator algebra at leading order in $1/N$. In this case,  \eqref{condo} can be solved for the part of $| \Psi_g \rangle$ that is sensitive to the action of the operator $\O$, obtaining \cite{Louko:1998hc}
 \be
|\Psi_g \rangle \sim \prod_{\om,m} \exp \left(\a_{\om,m} \, (-1)^m \O_{\om, m}^\dag \O_{\om, -m}^\dag \right) | 0 \rangle, \label{sentpsig}
\ee
where $| 0 \rangle$ is the analogue of the Rindler/Schwarzschild vacuum for the geon, satisfying $\O_{\om,m}|0 \rangle =0,\,  \forall \om>0$ and  $ \a_{\om,m}^{-1}  =2 G_\b(\om,m) \sinh \b \om/2 $.
This expression again displays the maximal entanglement between left- and right-movers, now valid in any CFT at large $N$. As explained in \cite{Louko:1998dj}, one can introduce a basis of wavepacket operators localised in time and space to turn \eqref{sentpsig} into an expression   that exhibits maximal entanglement between an operator  localized at time $t$ and position $\phi$  and one   localized at time $-t$ and position $\phi + \pi$. Tracing over the early-time modes, one obtains a thermal density matrix for the operators localized at late times. This gives a purely CFT derivation, at least in a large $N$ CFT, of the late-time thermality of the geon state from the point of view of generalized free field operators.

\subsection{Support of the geon state at high energies}
\label{deg}

We can also use the path integral expression for the geon state to characterise its support in an energy basis. The first interesting observation is that its support is not concentrated at a given energy. This makes our example different from many discussions of pure states in AdS/CFT, which focus on energy eigenstates. As noted in \cite{Harlow:2014yoa}, the latter are in fact not  well-approximated by a thermal ensemble as in \eqref{tcond}, but rather by the microcanonical ensemble. The geon state, by contrast, is by construction well-approximated by  a thermal ensemble for the late-time observables in $\mathcal A$. 

More concretely, we will show that the geon state has a Cardy-like growth of the spectral density with energy. This is clear for the free boson CFT from the expression \eqref{exprfbcc}: the cross-cap state has equal support on all states with equal left and right-moving quantum numbers, and the number of such states grows exponentially with energy. 

For a general CFT, we can derive a similar exponential growth at high energies by an exchange of two descriptions of the Euclidean geometry, analogous to the modular transformation of the torus in the thermal case.\footnote{We thank Alex Maloney for discussion on this point.} The cross-cap condition \eqref{vircc} implies that the cross-cap state is a sum over states with equal left- and right-moving energy ($L_0 = \bar L_0$), so we can write it as 
\begin{equation}  
| C\rangle = \sum_{i, m_i} C_{i,m_i} |i,m_i \rangle_L |i,m_i \rangle_R,
\end{equation}
where for each $i$, $|i, m_i \rangle_L$, ($|i, m_i \rangle_R$) is a basis of the left-moving (right-moving) states with $L_0 = \bar L_0 = E_i/2$ (so that the state $|i \rangle_L |i \rangle_R$ has energy $E = L_0 + \bar L_0 = E_i$). We can choose the basis such that the cross-cap state is diagonal, but the relative size of the contribution from different states is not fixed by \eqref{vircc}. Then, for the geon 
\begin{equation} \label{gstate2} 
|\Psi_g\rangle = \sum_{i, m_i}  e^{-\beta E_i/4} C_{i,m_i} |i,m_i \rangle_L |i,m_i \rangle_R.
\end{equation}
To determine the behaviour of the coefficients $C_{i, m_i}$, consider the CFT partition function on the Klein bottle, which is obtained by sewing together two M\"obius strips, and is given by 
\begin{equation}  \label{klein1} 
Z_K = \langle \Theta C | e^{-\beta H/2} | C \rangle  =  \sum_{E} e^{-\beta E/2} d_C(E_i).
\end{equation}
Here $\Theta$ is the CPT operator (for details, see \cite{Blumenhagen:2009zz}), and we defined the density of the support of the cross-cap state in energy,
\begin{equation} 
d_C(E_i) = \sum_{m_i} |C_{i,m_i}|^2.
\end{equation}
This is analogous to the density of states factor in the usual expression for the partition function on the torus. 

\begin{figure}[htp]
\centering 
\includegraphics[width=9cm]{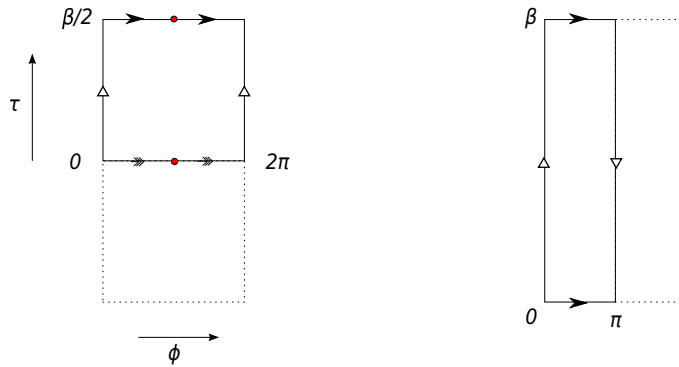}
\caption{The Klein bottle can be thought of as the quotient of a rectangular torus by the $\mathbb Z_2$ action $(\tau, \phi) \sim (-\tau, \phi+\pi)$. There are two natural fundamental regions for this identifications. In (a), we have a representation as the propagation between two cross-caps. In (b), we have the alternative representation with an orientation-reversing identification of the two sides.} \label{kb}
\end{figure}

 The Klein bottle can be represented as a rectangle in the $(\tau,\phi)$ plane, of height $\beta/2$ and width $2\pi$, with a periodic identification $\phi \sim \phi +2 \pi$ for $\tau \neq \{0, \beta/2\}$, and the cross-cap identification $\phi \sim \phi + \pi$ at $\tau = \{ 0, \beta/2\}$. One can alternatively represent the Klein bottle as a rectangle in the $(\tau,\phi)$ plane of height $\beta$ and width $\pi$, with the periodic identification $\tau\sim \tau + \beta$, and $(\tau, 0) \sim (-\tau ,\pi)$. This is an alternative fundamental region for the same identifications. The two are illustrated  in figure \ref{kb}. 
 
 In this alternative picture, consider interchanging the interpretation of $(\tau,\phi)$, so that $\phi = \tau'$ becomes the Euclidean time coordinate. After a conformal transformation, we obtain a rectangular region in the $(\tau', \phi')$ plane of height $2 \pi^2/\beta$ and width $2 \pi$, with the identifications $\phi' \sim \phi' + 2\pi$, $(0, \phi') \sim (2\pi^2 /\beta, -\phi')$. In this description, the Klein bottle partition function is 
\begin{equation} \label{klein2} 
Z_K = \mbox {Tr}( e^{-\frac{2 \pi^2 H}{\beta}} P),  
\end{equation}
where $P$ is the parity operator implementing the reversal of $\phi'$. 

This alternative description allows us to derive an analogue of the Cardy formula for the high-energy behaviour of the support $d_C(E)$. Consider high temperatures, so $\beta \ll 1$.  In \eqref{klein2},  the lowest-energy state will dominate in the trace, and thus
\begin{equation}
Z_K \approx e^{-\frac{2 \pi^2 E_0}{\beta}} \approx e^{\frac{2\pi^2 c}{12 \beta}} , 
\end{equation}
assuming the ground state is parity invariant.  This implies that the high energy asymptotics of $d_C(E)$ will be given by a Cardy-like formula, 
\begin{equation} \label{cardy}
d_C(E) \sim e^{\pi  \sqrt{ \frac{c E}{3}}}.
\end{equation}
There is a factor of two difference here from the usual Cardy formula for a finite-temperature CFT, where the degeneracy $d(E) \sim  e^{2\pi  \sqrt{ \frac{c E}{3}}}$. Thus, while the growth of the degeneracy at large temperatures/energies is qualitatively similar, the growth for the geon state is slower than for the thermal ensemble. 

This might seem surprising, but it is actually necessary for the relation between the average energy and $\beta$ for the geon state to agree with that in the thermal ensemble: the factor of 2 slower growth in the support cancels with the fact that the partition function \eqref{klein1} involves $e^{-\beta E/2}$ rather than the $e^{-\beta E}$ in the thermal partition function, so 
\be
\langle \Psi_g | H | \Psi_g \rangle = - 2 \p_\b Z_K 
= \frac{\pi^2 c}{3 \beta^2}
\ee
in agreement with the thermal result and also with the mass of the geon black hole as a function of $\beta$,
which  is  given by the BTZ formula. At late times, the geon black hole satisfies the first law with the same temperature, energy and entropy as the BTZ black hole; the entropy is given by the cross-sectional area of the horizon in the bulk at late times, which is the same as in BTZ. 

The entropy of the thermal ensemble approximating the geon state at late times provides the usual notion of coarse-grained entropy for this state. One can also associate to the geon state a different notion of entropy, by considering the dimension of the space of energy eigenstates on which it has significant support. The slower growth \eqref{cardy} of the support at large energies implies this will be half of the thermal entropy, matching the area of the horizon at $t=0$, which is \emph{half} of that at $t \neq 0$. This is also the area of the horizon in the Euclidean section, which was similarly related to an entropy in \cite{Krasnov:2003ye}. This mismatch between the two notions of entropy is not a contradiction: the geon is far from thermality at $t=0$, as discussed in section \ref{therm}, and thus thermodynamic notions of entropy  are not applicable at this time. Since it has a geometrical realisation, it would be interesting to  better understand the role of this alternative notion of entropy.

\subsection{Other boundary states}

The path integral construction reviewed above can be used to construct other  pure CFT states with a simple, analytic gravitational dual. For example, the state dual to the $J$ quotient of BTZ  that we discussed at the end of section \ref{defgeon} is obtained by evolving a Cardy boundary state $| B \rangle$ by $\b/4$ in Euclidean time  \cite{Maldacena:2001kr}
\be
| \Psi_B \rangle = e^{-\b H/4} | B \rangle. \label{bndst}
\ee
The reflection \eqref{geonq} imposes a Neumann boundary condition on the fields. In the free boson CFT for example, the  boundary state at $t=0$ satisfies
\be
(j_n + \bar j_{-n}) | B \rangle =0,
\ee
with solution (see e.g. \cite{Blumenhagen:2009zz})
\be
|B \rangle = \exp \left(-\sum_{k=1}^\infty \frac{1}{k}\, j_{-k} \, \bar{j}_{-k}\right) | 0 \rangle = \sum_{\vec{m}}| \vec{m} \rangle \otimes \overline{| U_b \vec{m} } \rangle, \label{exprbst}
\ee
where  $U_b$ is an anti-unitary operator that satisfies $U_b \, j_n \, U_b{}^{-1} = - j_n$, and we
 have again used the notation defined in \eqref{defvecm}. Thus, the Cardy boundary state  differs  from the crosscap state by a relative phase rotation between left-movers and right-movers. 

Since $|B \rangle$ is defined via the  identification $ t \sim -t$, similar arguments to those we used for the geon show that  $|\Psi_B \rangle$ satisfies  a relation directly analogous to \eqref{condo}
\be
e^{-\b H/2} \, \bar{\O}^\dag(-t,\phi)\, e^{\b H/2} | \Psi_B \rangle = \O(t,\phi) | \Psi_B \rangle .\label{condop}
\ee
When  $\O$ is a generalised free field operator, this relation is again the same as that satisfied by the analogue of the Hartle-Hawking state of the one-sided black hole dual to \eqref{bndst}.

\section{Mirror operators in the $\rp2$ geon geometry}
\label{mirror}

In this section, we  discuss the reconstruction of bulk operators in the $\rp2$ geon in terms of  boundary ones. We start by  reviewing the construction of local  bulk operators  in  BTZ in terms of smeared integrals of generalized free field operators. There is a well-known technical problem in this construction (see e.g. \cite{Leichenauer:2013kaa}), which is that when the bulk point approaches the horizon of the black hole, the integral over boundary operators spreads over all boundary times. Since later, in the geon case, we need to restrict to the algebra of late-time observables, this problem must be dealt with. We do this by considering  wavepackets of bulk excitations, which we show  are approximately localized in time from the boundary point of view. 

We then turn to the geon, and give an expression for the smeared bulk field operators inside the horizon. We use it to show that mirror operators in this case are just early-time local operators. This argument is most easily made in the high-frequency limit, where it reduces to a simple ray-tracing construction. 

\subsection{Local and smeared bulk operators in BTZ}
\label{btzsmear}

The construction of bulk operators outside a black hole's horizon in terms of CFT data has been investigated for some time, starting with the pioneering work of \cite{Banks:1998dd,Balasubramanian:1999ri,Bena:1999jv}. There is a well-developed proposal for operators in pure AdS \cite{Hamilton:2005ju,Hamilton:2006az} and BTZ \cite{Hamilton:2006fh}, where a local bulk operator is constructed in terms of boundary operators in a compact region,  at the price of complexifying the boundary. The authors of \cite{Papadodimas:2012aq} take a somewhat different approach, more in the spirit of the original work in \cite{Banks:1998dd}, working in momentum space and reconstructing bulk momentum modes in terms of boundary momentum modes. 

For an eternal black hole (in our case, BTZ), the proposal is that the bulk field at a point outside the horizon (in region I) can be reconstructed as \cite{Papadodimas:2012aq}
\begin{equation} \label{bulkout}
\Phi^I(t, \phi, r) =  \sum_{m \in \mathbb{Z}} \frac{1}{(2\pi)^2} \int_{\omega >0} d \omega \, \left[ \mathcal O_{\omega, m}\, \varphi_{\om, m} (t,r,\phi) + \mathcal O_{\omega, m}^\dag\, \varphi_{\om, m}^\star (t,r,\phi)  \right], 
\end{equation}
where $\mathcal O_{\omega, m}$ are the Fourier modes of the corresponding CFT operator
 on the asymptotic boundary of the right exterior region
\be
\O_{\omega, m} = \int dt d\phi  \, e^{i \om t - i m \phi} \,  \O(t,\phi)
\ee
and $\varphi_{\om,m}$ is a plane wave basis of normalizable solutions to the scalar field equation in BTZ
\be
\Box \varphi_{\om,m} = M^2 \varphi_{\om,m} \;, \;\;\;\;\; \varphi_{\om,m}(t,r,\phi) = e^{-i \om t + i m \phi} f_{\om,m} (r). \label{waveqn}
\ee
The explicit expression for $f_{\om,m}(r)$ is given in appendix \ref{expl}. It satisfies 
\be
f_{\om,m}(r) = f^\star_{\om,m}(r)\;,\;\;\;\;\;f_{\om,m}(r) = f_{\om,-m}(r)
\ee
and as $r \r \infty$ it behaves as $f_{\om,m} \sim  r^{-\Delta}$.  For a point inside the black hole ($r<r_+$), by contrast, the proposed reconstruction involves operators on both boundaries,
\begin{equation} \label{bulkin}
\Phi^{II}(t, \phi, r) =  \sum_m \frac{1}{(2\pi)^2}\int_{\omega >0} d \omega \left[ \mathcal O_{\omega, m} \, \chi^{(+)}_{\omega, m}(t,r,\phi)  + \tilde{ \mathcal  O}^\dag_{\omega, m} \, \chi^{(-)}_{\omega, m}(t,r,\phi) +\mbox{h.c.} \right]    
\end{equation}
where $\tilde \O_{\omega, m}$
are the Fourier modes of the  operator corresponding to $\Phi$ in the CFT on the left asymptotic boundary, which are defined with the opposite signs to take into account the different direction of the time-translation on the left boundary, 
\be
\tilde \O_{\omega, m} = \int dt d\phi  \, e^{-i \om t + i m \phi} \,  \tilde \O(t,\phi)
\ee
h.c. denotes the hermitian conjugate, and
\be
\chi^{(\pm)}_{\omega, m} (t,\phi,r) =  e^{-i \omega t + i m \phi} g^{(\mp)}_{\omega, m}(r)
\ee 
are two linearly independent plane wave mode solutions in region II. This basis of solutions is chosen so that as $r \to r_+$
\be 
g^{(\pm)}_{\om,m} \sim e^{\pm i \om r_\star} \; , \;\;\;\;\; r_\star = \frac{\ell^2}{2 r_+} \ln \left| \frac{r-r_+}{r+ r_+}  \right| \label{defrst}
\ee
where $r_\star$, defined above,  is the tortoise radial coordinate. Thus, our chosen mode functions consist of  a ``left-moving'' mode solution $\chi^{(+)}_{\om,m}$, which enters region II from region I, and a ``right-moving'' mode solution  $\chi^{(-)}_{\om,m}$, which enters region II from region III. It is not hard to check, using the explicit expressions given in appendix\footnote{The functions $g^{(\pm)}_{\om,m}$ are related to the functions $f^{(\pm)}_{\om,m}$ defined in  appendix \ref{expl} by a simple rescaling.} \ref{expl}, that $g^{(\pm)}_{\om,m}$ are complex conjugates of each other, and moreover that $g^{(\pm)}_{\om,m} =g^{(\pm)}_{\om,-m} $. 

For generic points outside or inside the horizon, it is possible to invert the Fourier transform defining $\mathcal O_{\omega, m}$ to rewrite the expression  for $\Phi(t,\phi,r)$  in terms of an integral over $\mathcal O(t', \phi')$, as in \cite{Hamilton:2005ju,Hamilton:2006az}. For example, the field in region I can be written as
\be
\Phi^I(t,r,\phi) = \int dt' d\phi' K(t-t',r,\phi-\phi') \O (t',\phi'),
\ee
where the integration kernel $K(t-t',r,\phi-\phi')$ is the Fourier transform of $f_{\om,m}(r)$ with respect to $\om, m$
\be
K (t-t',r,\phi-\phi') = \sum_m \frac{2}{(2\pi)^2}\int_{\omega >0} d \omega \,   \cos \left[\om (t-t')-m (\phi-\phi') \right] \, f_{\om,m}(r).
\ee
When the bulk point of interest is close to the boundary, $K(t-t',r,\phi-\phi')$ has support over a range of boundary times $\Delta t'$ of order $\ell^2/r$, centered at $t$. However, as we approach the horizon, this integral delocalises, spreading over all boundary times. In Fourier space, this delocalization is represented by a divergence in the integrand\footnote{Nevertheless, as shown in \cite{Papadodimas:2012aq}, this divergence does not affect physical correlators of the operators in question.} as $\om \r 0$. 
 Similarly, as emphasized in \cite{Leichenauer:2013kaa}, the  sum over momenta $m$ is divergent in any black hole background, and needs to be regularized. These issues can be addressed by considering wave packets, rather than point localised bulk field operators. For our purposes, approximate boundary localization of the bulk observables is necessary, since in the geon the algebra of observables is restricted to boundary operators supported at $t > t_*$.

We therefore consider the relation between a bulk wave packet built from the field operator $\Phi(t,\phi,r)$ and boundary wave packets built from the operators $\mathcal O(t,\phi)$. Note that we will only be smearing  in the $t, \phi$ directions; the dependence of the bulk modes on the $r$ direction is  determined dynamically by the equations of motion. 

We construct wave packets in a standard way (see e.g. \cite{Takagi:1986kn}), by defining
\begin{equation} \label{boxsm}
\tilde{\xi} _{\omega_0 t_0}(\omega) = \left\{ \begin{array}{ll} \frac{1}{\sqrt{\epsilon}} \, e^{ i \omega t_0} & \mbox{ for } \omega \in (\omega_0 - \frac{1}{2} \, \epsilon, \omega_0 + \frac{1}{2} \, \epsilon), \\ 0 & \mbox{ otherwise.} \end{array} \right. 
\end{equation}
These wavepackets are centered about $\omega_0$ with width\footnote{In \cite{Louko:1998dj,Takagi:1986kn}, it is required that $\om_0 = (p+ \half) \e$ and $t_0 = 2\pi q/\e$ with $p,q \in \mathbb{Z}$; this integrality property is necessary for proving orthonormality and completeness of the set of wavepackets. Since we will not make explicit use of these properties here, we have settled for the more physically transparent notation above. } $\epsilon$. The Fourier transforms
\be
 \xi _{\omega_0 t_0}(t) = \frac{1}{2\pi} \int d\om \, e^{-i \om t} \, \tilde{\xi}_{\om_0,t_0} (\om)  
\ee 
 (whose explicit expression we will not need) are cardinal sinus functions supported about $t_0$ with width $2 \pi \epsilon^{-1}$. We will generally consider $\epsilon$ of order one, but take large $\omega_0, t_0$. We can similarly introduce wavepackets localized in momentum space around $m_0$, with an angular central position $\phi_0$ 
\be \label{etasm}
\tilde{\eta} _{m_0,\phi_0}(m) = \left\{ \begin{array}{ll} \frac{1}{\sqrt{\e}}\, e^{- i m \phi_0} & \mbox{ for }m \in (m_0 - \frac{1}{2} \,\e, m_0 + \frac{1}{2} \, \e), \\ 0 & \mbox{ otherwise.} \end{array} \right.
\ee
For simplicity, we have taken the smearing in momentum and frequency to be the same, but it is easy to have them different. We can now smear the bulk field operator $\Phi(t,r,\phi)$, by convolving with $\xi_{\om_0,t_0}(t)$ and $\eta_{m_0, \phi_0}(\phi)$. It is useful to first construct a wavepacket consisting purely of annihilation operators
\begin{equation} \label{phiplus}
\Phi^{+}_{\omega_0,m_0}( t_0,r, \phi_0) = \int dt\, d\phi\, \xi^\star_{\omega_0 t_0}(t)\, \eta^\star_{m_0,\phi_0} (\phi) \, \Phi^{+}(t,r,\phi),
\end{equation}
where $\Phi^{+}$ contains only the first term in  \eqref{bulkout}. At the end, we shall add to \eqref{phiplus} its hermitean conjugate. The resulting hermitean operator is localized  in position space around the bulk point $(t_0, r, x_0)$ with width of order $2\pi \e^{-1}$, and  is box normalized in momentum space\footnote{Note that our smearing differs slightly from that of \cite{Louko:1998dj}, who considered instead    the wavefunctions  
\begin{equation*}
\bar{\varphi}_{\om_0, t_0; m_0 ,\phi_0} (t,r,\phi) = \sum_m \frac{1}{(2\pi)^2} \int d\om \, \tilde \xi_{\om_0, t_0} (\om) \, \tilde \eta_{m_0, \phi_0} (m) \, \varphi_{\om,m} (t,r,\phi) ,
\end{equation*}
which were then convolved with the smeared creation-annihilation operators. While this method also yields a bulk field operator that is localized in position space, its bulk localization is under less control than that defined above. The difference is that our procedure fixes the amount of smearing around the bulk point $(t_0,r,\phi_0)$, and lets the spread on the boundary  to be determined by  the Fourier transform of $f_{\om,m}(r)$; the smearing used in \cite{Louko:1998dj} fixes instead the spread of the operator on the boundary, and lets the one in the bulk be determined by the Fourier transform of $f_{\om,m}(r)$.} around $(\om_0,m_0)$, with width of order $\e$. 

Note that this smearing has relatively strong tails in the wave packets; our smeared operators are analytic functions in position space, so the solution in the interior can be reconstructed given its values in an open neighbourhood of the boundary.  However, for the physics we plan to analyse, this is just a technical issue. We expect qualitatively similar conclusions would be obtained if we used some other smearing (e.g. Gaussian). 

\begin{figure}
\centering
\includegraphics[width=11 cm]{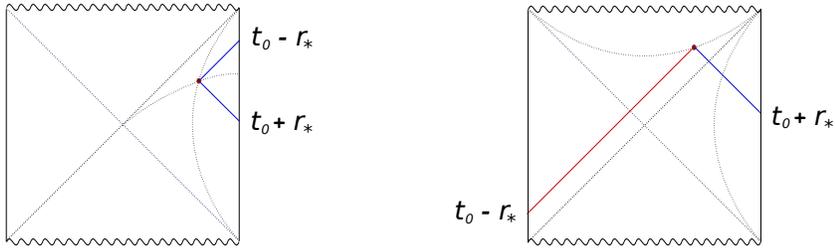}

\caption{Support of high-frequency modes on the boundary of the eternal BTZ spacetime for a mode outside the horizon (left) and one behind the horizon (right). The entanglement of the modes on the left and right boundaries in the thermofield double state ensures the smoothness of the bulk field operator as we cross the horizon.} \label{btzsmfig}
\end{figure}

Plugging in the expression  \eqref{bulkout} into \eqref{phiplus} and replacing $\O_{\om,m}$ by its Fourier transform, one finds 
\begin{equation} 
\Phi^{I}_{\omega_0, m_0} (t_0,r,\phi_0) = \int dt d \phi \, K^{I}_{\om_0,m_0} (t_0-t,r,\phi_0-\phi) \O(t,\phi), \label{wavep}
\end{equation}
where we have defined the smeared kernel in region $I$,  $K^{I}_{\om_0,m_0} $ via
\bea
K^I_{\om_0,m_0} (t_0-t,r,\phi_0-\phi)& = & \sum_m \frac{1}{(2\pi)^2}  \int d\om \, \left[  e^{i \om t- i m \phi} \tilde{\xi}^\star_{\om_0,t_0} (\om) \tilde \eta^\star_{m_0,\phi_0}(m) + \mbox{h.c.} \right] f_{\om,m}(r) \non\\
& &\hspace{-2.5 cm} = \sum_{m=m_0 - \half \e}^{m_0+\half \e} \frac{2}{(2\pi)^2 \e}  \int_{\om_0 - \half \e}^{\om_0 + \half \e} d\om \, \cos \left[ \om (t_0-t) - m (\phi_0-\phi)\right] f_{\om,m} (r).\label{smki}
\eea
We can understand the relation between the bulk and boundary operators in this wave packet very simply if we consider the case of high-frequency wave packets, where  the central frequency $\omega_0$ is taken to be large compared to the spacetime curvature scale $\ell^{-1}$ and the mass $M$ (if any) of the bulk field $\Phi$.  Then, in the radial direction the effective scattering potential is unimportant away from the boundary (for $r < \ell^2 \omega_0$) and the mode solutions $f_{\omega, m}(r)$ for $\omega \approx \omega_0$ take a wavelike form  $f_{\omega,m}(r) \sim r^{-\half}\cos( \omega r_* + \d_{\om,m})$ \eqref{appsolf}, where $r_*$ is the tortoise coordinate in the BTZ black hole defined in \eqref{defrst}. As shown in appendix \ref{smearing}, the momentum integral in \eqref{smki}  gives a result  concentrated around $t = t_0 \pm r_*$. This can be understood as the result of the propagation at these high frequencies being described by geometric optics (ray tracing), as illustrated in figure \ref{btzsmfig}. 

Within this wave packet picture, it is easy to understand what happens as we approach the horizons. If we approach the future horizon, then $t_0  \to \infty$, $r_* \to -\infty$ with $t_0 + r_\star$ approximately constant. In this limit, the contribution localised on the boundary at $t_0 + r_*$ remains at a finite position, while the contribution localised at $t_0 - r_*$ is sent off to  infinity. That is, the operator in the field theory corresponding to a bulk mode near the horizon has a contribution from arbitrarily late times.

We can consider the same wave packet construction for the operators inside the horizon, yielding 
\be
\Phi^{II}_0 (t_0,r,\phi_0) = \int dt d \phi \, \left[ K^{(+)}_0 (t_0-t,r,\phi_0-\phi) \, \O(t,\phi) + K^{(-)}_0 (t_0-t,r,\phi_0-\phi) \, \tilde \O(t,\phi)  \right]
\ee
where, for simplicity, we have replaced the subscript ``$\om_0,m_0$'' by just ``$0$'' and have defined 
\bea
K^{(\pm)}_{0} (t_0-t,r,\phi_0-\phi)& = & \sum_m \frac{1}{(2\pi)^2}  \int d\om \, \left[  e^{i \om t- i m \phi} \, \tilde{\xi}^\star_{\om_0,t_0} (\om) \, \tilde \eta^\star_{m_0,\phi_0}(m) \, g^{(\mp)}_{\om,m}(r) + \mbox{h.c.} \right]   \non\\
& &\hspace{-2.4 cm} = \sum_{m=m_0 - \half \e}^{m_0+\half \e} \frac{1}{(2\pi)^2 \e}  \int_{\om_0 - \half \e}^{\om_0 + \half \e} d\om \, \left[ e^{-i \om (t_0-t) + i m (\phi_0-\phi)} g^{(\mp)}_{\om,m} (r) + \mbox{h.c.} \right].
\eea
For high frequencies, $g^{(\pm)}_{\omega, m}(r) \sim e^{\pm i \omega r_*}$, so that $K_0^{(+)}$, which multiplies  the operator $\mathcal O$ on the right boundary, has support at $t= t_0 + r_*$, while $K_0^{(-)}$, which accompanies the operator $\tilde{\mathcal O}$ on the left boundary,  is  supported at $t = t_0 - r_*$, as illustrated in figure \ref{btzsmfig}. Hence, again, if we approach the horizon on the right, the contribution from the right boundary remains at finite position, but the contribution on the left boundary is going off  to the infinite past.

\subsection{Local and smeared bulk operators in the geon geometry}

We now consider the reconstruction of bulk field operators in the geon geometry.  In the region outside the black hole, the mode solutions are the same as in BTZ, so the expression for the bulk field $\Phi^I(t,r,\phi)$ in terms of boundary operators is still given by \eqref{bulkout}.

Inside the horizon, region II in the geon can be described by the same coordinates as  region II in BTZ, but with the time coordinate restricted to $t >0$. The solutions of the equations of motion in this region should be invariant under $(t,\phi) \r (-t, \phi+\pi)$, and can be constructed from the solution in region II of BTZ by the method of images. There will then not be independent left- and right-moving solutions, but a single one, determined by this boundary condition. Thus, in region II we have
\begin{equation} \label{geonin}
\Phi^{II}_g (t,r, \phi) =  \sum_m \frac{1}{(2\pi)^2} \int_{\omega >0} d \omega [ \mathcal O_{\omega, m} (e^{-i \omega t + i m \phi} + (-1)^m e^{i \omega t + i m \phi}) g^{(-)}_{\omega, m}(r) + \mbox{h.c.} ],
\end{equation}
where $g^{(-)}_{\omega, m}(r)$ is the same function as in the BTZ case. This expression for $\Phi^{II}_g$ can be obtained directly from the one in BTZ, \eqref{bulkin}, by making the \emph{formal} replacement  
\begin{equation}
\label{formrep}
\tilde{\O}_{\om,m}\;\; \longrightarrow \;\; (-1)^m \O_{\om,-m} 
\end{equation}
at the level of the Fourier modes. Note that even though  \eqref{geonin} is only defined for $t>0$, the analytic continuation of this expression to $t<0$ yields a bulk field satisfying the boundary condition $\Phi^{II}(t,r,\phi) = \Phi^{II}(-t,r,\phi+\pi)$. This point of view will be useful in section \ref{deviations}.

\begin{figure}
\begin{center}
\includegraphics[width=11.5cm]{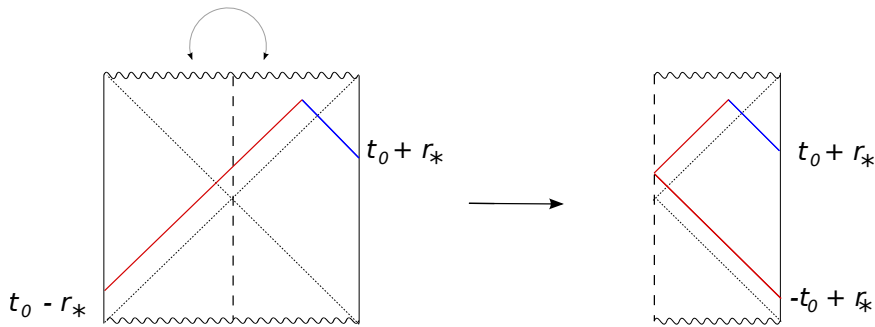}
\caption{The support of a high-frequency mode behind the horizon in the geon, as understood from the quotient construction from BTZ.}
\label{geonwp}
\end{center}
\end{figure}

We can pass to wave-packets as in BTZ, with the result that the CFT operator corresponding to a smeared bulk field   outside the horizon is given by \eqref{wavep}, just as before. For a point behind the horizon, the CFT operator corresponding to the wave packet is
\be \label{geoninwp}
\Phi^{II}_{ \omega_0,m_0} (t_0,r,\phi_0) = \int dt d \phi \, \left[ K^{(+)}_0 (t_0-t,r,\phi_0-\phi) + K^{(-)}_0 (t_0+t,r,\phi_0-\phi-\pi)    \right] \O(t,\phi) .
\ee
If we consider high frequencies, the operator wavepacket  for a point outside the horizon has boundary support   concentrated at $t = t_0 \pm r_*$, as in BTZ, but  for a point inside the horizon, its  boundary support is concentrated at $t = t_0 + r_*$, $t = -t_0 + r_*$. This can again be understood in terms of geometric optics: the right-moving part of the wave packet inside the horizon reflects off the geon identification and intersects the boundary at $r_\star - t_0$, as illustrated in figure \ref{geonwp}. 

\subsection{Identification of the mirror operators}
\label{mid}

The mirror operators are defined to be the solutions of \eqref{conj}, \eqref{comm} for the geon state in the dual CFT. They  can also be found by comparing the expression for the bulk field $\Phi$ inside the horizon of the geon at  large $t$ with the corresponding expression in BTZ. Comparing  \eqref{geonin} to \eqref{bulkin}, we can identify the second term with the contribution from the mirror operators, and thus 
\begin{equation} \label{mirid}
\tilde{\O}^g_{\omega,m} = (-1)^m \O_{\omega,- m}\;\;\;\;\; \Rightarrow \;\;\;\;\;\tilde{\O}_g(t, \phi) = \O(-t, \phi+\pi) .
\end{equation}
Thus, the mirror of an operator at late times is the same operator, acting at early times.
We emphasize that this derivation of $\tilde{\O}_g$ is only expected to hold for large $t$. Having made this identification, we would like to verify as far as possible that these mirror operators satisfy \eqref{conj}, \eqref{comm}. 

The first condition \eqref{conj} is simply that $\tilde \O_g$  is maximally entangled with $\O$ in the geon state.  This condition was verified from the CFT point of view in  section \ref{entstr}. Nevertheless, since we did not have a complete proof of \eqref{condo}  in a general CFT, we will now review an alternative bulk argument for its validity at leading order in $1/N$ for generalised free field operators in a large $N$ CFT.

\begin{figure}[ht]
\centering
\includegraphics[height=3.4cm]{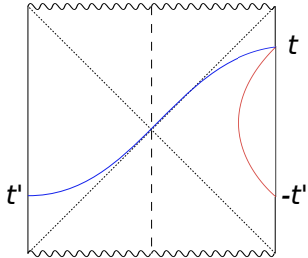}
\caption{Contributing geodesics to the two-point function of $\O(t)$ and $\O(-t')$ in the geon using the method of images. The geodesic between the two boundaries in BTZ represents a contribution to the correlator which depends only on $t-t'$, signalling the entanglement of these modes. By contrast, the contribution to the correlator from geodesics linking points in the same boundary in BTZ is exponentially small as $t+t'$ becomes large. }
\label{geongeod}
\end{figure}

The geon state is analogous to the Hartle-Hawking vacuum in BTZ,  and can be constructed as usual by demanding that  it  be annihilated by all modes $d^g_{\om,m}$ of positive frequency with respect to an infalling observer's time \cite{Louko:1998dj}
\be
 d^g_{\om,m} | \Psi_g \rangle = \frac{e^{\b \om/4}}{\sqrt{2 \sinh \frac{ \b \om}{2} \, G_\b(\om,m) (1-e^{-\b \om})}}  \left( \O_{\om,m} - e^{-\b \om/2} (-1)^m \O^\dag_{\om,-m}  \right) | \Psi_g \rangle =0 ,\label{gannih}
\ee
where we have used the appropriate Bogoliubov transformation to rewrite this condition in
 terms of the asymptotic (Rindler/Schwarzschild) creation-annihilation operators $\O^{(\dag)}_{\om,m}$.
Note that this is the same equation as we would have obtained by starting with the equation for the Hartle-Hawking state for BTZ and making the replacement \eqref{formrep}. Passing to position space, we find that the geon state satisfies
\bea
e^{-\b H/2} \O^\dag(t,\phi) e^{\b H/2}| \Psi_g \rangle &=& \sum_m \frac{1}{(2\pi)^2} \int_{\om>0} d\om \left( e^{i\om t-i m \phi} e^{-\b \om/2} \O^\dag_{\om,m}   +  e^{-i\om t+i m \phi} e^{\b \om/2} \O_{\om,m} \right) | \Psi_g \rangle \non \\
&=& \sum_m \frac{1}{(2\pi)^2} \int_{\om>0} d\om \left( e^{i\om t-i m \phi} (-1)^m \O_{\om,-m} + \mbox{h.c.} \right)  | \Psi_g \rangle \non \\
&=& \O(-t,\phi + \pi)|  \Psi_g \rangle, \label{ftmir}
\eea
which is exactly the same relation that we argued in section \ref{pathintg} should hold from the CFT point of view.\footnote{Assuming the validity of the CFT derivation, the relation \eqref{ftmir} does not receive any correction in $1/N$, perturbative or non-perturbative. Thus, it may provide an interesting departure point for studying the effect of such corrections to the PR proposal.} This entanglement is inherited, via the quotient construction,  from that between the degrees of freedom on the right and on the left in the eternal BTZ black hole. It can be understood pictorially by considering the geodesics connecting points on the two boundaries of BTZ, as illustrated in figure \ref{geongeod}. The BTZ time-translation invariance implies that the length of these geodesics depends only on $t-t'$, so the corresponding contribution to correlators is large for $t' \approx t$. By the method of images, the same is true in the geon. 

\bigskip

The second condition \eqref{comm} is only expected to be satisfied for sufficiently large $t$. It is clear that the two operators $\O(t)$, $\O(-t')$  are timelike separated, so their commutator does not vanish as an operator statement (unlike in the BTZ case, where the commutator of operators on the different boundaries vanishes identically). We have not been able to calculate it directly from the CFT side, but we can easily calculate its expectation value holographically, using the method of images formula \eqref{twopt}. As noted above, the contribution from the second term in \eqref{twopt} is independent of $t+t'$. However, this part makes no contribution to the commutator in the bulk, as it corresponds to  spacelike separated points in BTZ. The contribution from the first term comes from points on a single boundary in BTZ, as illustrated in figure \ref{geongeod}. So it does contribute to the commutator, but this contribution becomes exponentially small at large $t+t'$ because of the quasinormal decay of bulk fields.\footnote{Note that PR require that the mirror operator satisfy \eqref{comm} exactly, whereas our proposed mirror operator only satisfies it inside correlation functions in the state $|\Psi \rangle$ and only up to terms exponentially supressed in the black hole entropy.
 To obtain a mirror operator satisfying \eqref{comm} exactly might require us to add exponentially small corrections to these simple single-trace operators.} This argument can easily be extended to the expectation value of the commutator inside all low-point correlation functions, as in section \ref{therm}.\footnote{There is an important exception to this general argument, when  we take the operator $\mathcal O(t)$ to be the Hamiltonian (or any other conserved charge, if the dual CFT contains any). Because it is conserved, the commutator of mirror operators with the Hamiltonian will not vanish. This is a general exception, however, which applies to the mirror operator construction in any state and is discussed in \cite{Papadodimas:2013jku}. } 

Thus, we argue that the formula \eqref{mirid} for the CFT operator corresponding to a field mode inside the horizon of the geon provides an interesting explicit example of the PR construction, where the mirror operators of local single-trace operators at late times are simply local single-trace operators at early times. 
 
 \subsubsection*{Remarks} 
 \begin{itemize}
 \item  Note that if we picture the geon as a state  defined on the $t=0$  surface by analytic continuation from the Euclidean path integral, the simplicity of the mirror operators is only apparent. The operators $\mathcal O(-t)$ are actually precursors in the sense of \cite{Susskind:2013lpa,Susskind:2013aaa}, defined by evolving the state backwards in time to $-t$, acting with this local operator and then evolving forward in time to $t$. This folded time prescription defines a complicated non-local operator acting at $t=0$ (or $t>0$ if we apply further evolution). 

\item The simple picture of the mirror operators as early-time local operators is very special to the $\rp2$ geon. For the more general geons of \cite{Aminneborg:1997pz}, because of the shadow region behind the horizon, ray-tracing back the right-moving part of the interior mode  we reach the past singularity, rather than the asymptotic boundary. One way of understanding what makes the $\rp2$ geon so special is that in this case, any solution of the wave equation on the geon descends from an even solution of the wave equation on BTZ. While  more complicated single-exterior geons can also be understood as quotients of BTZ, due to fixed points of the quotients on the  BTZ boundary, there will be solutions to the wave equation on the geons which do not lift to solutions  on BTZ that satisfy the boundary conditions everywhere\footnote{We thank Ian Morrison for this comment.}.

\item For more general smooth collapse situations, ray-tracing will fail (as has been emphasised recently in e.g. \cite{Almheiri:2013hfa}), because one encounters a trans-Planckian problem if he tries to follow a mode inside the black hole at $t > t_*$ backwards in this way. This issue is evaded in the geon case because the blueshift as we follow the mode back along the future horizon is balanced out by a redshift as we follow it out along the past horizon to the asymptotic boundary, so that the description in terms of operators in the boundaries involves the same coordinate frequency as we started with. This cancellation is clearly delicate, as it requires the mode to re-emerge from the black hole exponentially close to the past horizon, and it will not generalise to other collapse scenarios. 

\item Let us note however that another simple example in which the ray-tracing argument is applicable is the $J$ quotient of BTZ, which has an orbifold singularity inside the horizon.  In this case, the analogue of the Hartle-Hawking like state satisfies 
\be
 \O_{\om,-m} | \Psi_B \rangle = e^{-\b \om/2} \O^\dag_{\om,m} | \Psi_B \rangle, \label{condbst}
\ee
which, via arguments identical to those we used in the geon, leads to an identification of the mirror operators as $\tilde \O_B(t,\phi) = \O(-t,\phi)$. This can again be understood as  ray-tracing a right-moving mode behind the horizon backwards towards the singularity at $X_2 =0$, off which it reflects  just like in the geon, but without the shift by $\pi$.  
\end{itemize}

\section{Modifications of the geon  state}
\label{deviations}

The central assumption of the PR construction is that the bulk geometry contains a black hole with a smooth, empty interior that looks like a patch of  the eternal black hole at late times. However, should such a prescription be applied indiscriminately to all states, it would lead to the so-called ``frozen vacuum" problem \cite{Bousso:2013ifa}: the procedure would always construct operators behaving like local fields in an empty interior, even though there exist states whose holographic dual does not have an empty interior. PR have argued that the equilibrium condition \eqref{tcond} will identify the states to which their prescription can be applied. Non-equilibrium states generated by acting on the equilibrium state $|\Psi \rangle$ with operators in $\mathcal A$ will not satisfy \eqref{tcond}, and would then be identified by their prescription as excitations falling into the black hole. States obtained by acting on $|\Psi \rangle$ with a mirror operator can also be detected as out of equilibrium, since the Hamiltonian (and possibly other conserved charges) does not commute with $\tilde{\mathcal O}_\Psi$  \cite{suvratstrings}. 

A more challenging example was proposed in \cite{Harlow:2014yoa}.  Consider a unitary operator built out of the mirror operators, which commutes with the Hamiltonian, for example
\begin{equation}
|\Psi' \rangle = e^{i \theta(\om)\tilde{\mathcal O}_\omega^\dagger \tilde{\mathcal O}_\omega} |\Psi \rangle. \label{psip}
\end{equation}
The operator in the exponential  is the number operator for a right-moving mode behind the horizon at late times. Acting with the above deformation is applying a phase rotation which disrupts the entanglement between modes just inside and just outside the horizon.
The action of such a unitary operator is not detectable by observables in $\mathcal A$, now including the Hamiltonian\footnote{This statement holds at strictly infinite $N$. The operator appearing in \eqref{psip} can be corrected order by order in $1/N$, such that it commutes with the Hamiltonian to arbitrarily good precision. }. Thus, if $| \Psi \rangle$ is an equilibrium state satisfying \eqref{tcond},  $|\Psi' \rangle$ will be as well.

Using the relation \eqref{conj} between the operators $\O$ in $\mathcal A$ and their mirrors,  \cite{Harlow:2014yoa} argued that \eqref{psip} could be rewritten, to arbitrarily good precision, as a unitary rotation constructed from the operators in $\mathcal A$,
\be
|\Psi' \rangle = e^{i \theta(\om) {\mathcal O}_\omega^\dagger {\mathcal O}_\omega} |\Psi \rangle .\label{psipp}
\ee
Such a state has been previously considered by \cite{Almheiri:2013hfa} and poses an important challenge to the PR construction, because now  $|\Psi' \rangle$ can be viewed either as an equilibrium state in itself, or as an excitation of the equilibrium state $|\Psi \rangle$. The two interpretations lead to different predictions for the mirror operators and, more importantly, for the experience of an infalling observer, indicating  an ambiguity in the PR proposal. 

Since the unitary phase rotation in \eqref{psip}, \eqref{psipp} changes the entanglement between modes inside and outside the horizon, it can be argued, by analogy with the Minkowski vacuum in Rindler quantization, that such states should not have a smooth horizon (see e.g. \cite{Harlow:2014yka}). There, changing the relative phase between the left- and right-moving Rindler mode pairs does not change the density matrix  (and thus the entanglement entropies) in either the left or right Rindler wedges, but it does lead to a state that is no longer the global Minkowski vacuum, but an excited state.  This is easy to see, as the new state is no longer annihilated by the Minkowski annihilation operators.  Moreover, this state is not smooth across the Rindler horizon if the modes whose relative phase we change are of high frequency. A similar argument can be applied near any Killing horizon, in particular that of a black hole. Of course, in the case of an emergent space-time (rather than a fixed background such as Minkowski space), the geometry can readjust itself in the new state so that the horizon is still smooth.

In the following, we would like to see whether we can  shed light on the issues raised by \cite{Almheiri:2013hfa,Harlow:2014yoa} by applying  unitary rotations of the type \eqref{psip},\eqref{psipp} to the geon state $|\Psi_g \rangle$.  We consider two different examples.

\subsection{A  special unitary rotation}

We have discussed two pure CFT states  that are expected to correspond to single-sided black holes in the bulk: the geon state, which we argued should satisfy property \eqref{condo}, and the state \eqref{bndst} dual to the $J$ quotient of BTZ,  expected to satisfy \eqref{condop}. At large $N$, and as far as observations involving the generalized free field operators $\O$ are concerned, these two states are related by a unitary rotation. This can easily be seen by solving the conditions  \eqref{condo}, \eqref{condop} to write these states as 
 \be
|\Psi_g \rangle \sim \prod_{\om,m} e^{\a_{\om,m} (-1)^m \, \O_{\om, m}^\dag \O_{\om, -m}^\dag} | 0 \rangle \;, \;\;\;\;\;\; 
|\Psi_B \rangle \sim \prod_{\om,m} e^{\a_{\om,m} \, \O_{\om, m}^\dag \O_{\om, -m}^\dag } | 0 \rangle.
\ee
While it is not clear whether these states are related by a unitary rotation at the level of the full Hilbert space,\footnote{In the free boson CFT, the crosscap state $|C \rangle$ and the boundary state $| B \rangle$ are related by a unitary rotation, as can be easily checked by comparing \eqref{exprfbcc} and \eqref{exprbst}. This implies that $|\Psi_g \rangle$ and $|\Psi_B\rangle$ are also related by it. For rational CFTs, the crosscap and boundary Ishibashi states are also related by a unitary rotation \cite{Blumenhagen:2009zz}; nevertheless, the expressions for $|C \rangle$ and $| B \rangle$ as linear combinations of Ishibashi states  involve also certain reflection coefficients, which do not appear to be  simply related. } the fact that they are related by a unitary rotation at large $N$ is sufficient for us to treat this as an example of the general issues discussed above. 

The state $|\Psi_B \rangle$ is not annihilated by the ``Unruh'' annihilation operators \eqref{gannih} in the geon so, according to  the arguments of  \cite{Almheiri:2013hfa}, its  horizon is not smooth. This is not what we find: the spacetimes dual to both $|\Psi_B \rangle$ and $| \Psi_g \rangle$ have a smooth horizon and an identical geometry in region I. The difference between the two states is encoded  only in the geometry behind the horizon, which is smooth in the geon case but has an orbifold singularity for the $J$ quotient of BTZ. Also,  the global properties of the two space-times are different, as the $J$ quotient of BTZ is orientable, while the $\rp2$ geon is not. 

The unitary rotation between these states has a special character: the change in the  entanglement pattern between $|\Psi_B \rangle$ and $| \Psi_g \rangle$  corresponds to a relative rotation by $\pi$ in the  $\phi$ direction of the entangled modes outside and inside the horizon. Changes in entanglement generated by similar symmetry transformations have been previously discussed in \cite{Maldacena:2013xja}, where they were also argued to preserve the smoothness of the horizon. Our example has the advantage that unlike in \cite{Maldacena:2013xja}, where the background geometry was fixed (Minkowski space) and there simply existed a \emph{choice} of quantization in which the horizon looked smooth, in our case we have a dynamical space-time, and we can explicitly see that the saddle-point geometry in the path integral re-adjusts itself to produce a smooth horizon.  

Thus, we provided an example where the correct geometry dual to a unitarily rotated state has a smooth horizon. The leading-order effect of the unitary rotation is to change the geometry in the deep interior, on the  $X_2=0$ line.

\subsection{Infinitesimal mode rotations}

The unitary transformation considered in the previous section acts on all modes of all fields simultaneously - it corresponds to a geometrical rotation. We will now consider states constructed by a rotation acting only on certain modes of the operators, as in \eqref{psip}-\eqref{psipp}. These provide a more typical set of examples, where the unitary transformation can no longer be given a simple geometrical interpretation.

We will consider a rotation acting on the following smeared operators
\be
\O_{\om_0,t_0} = \int d\om\, \tilde \xi^\star_{\om_0,t_0} (\om) \,\O_\om,
\ee
where $\tilde \xi_{\om_0,t_0}(\om)$ has been defined in \eqref{boxsm}. One can easily check that $\O_{\om_0,t_0}$ is  localized in momentum space around $\om_0$ with width $\e$, and in time around $t_0$, with width $2\pi/\e$.\footnote{Unlike before, here we may consider a large spread in time, so $\e$ can be small.}
Note that, in order to simplify the notation, we have completely suppressed the coordinate $\phi$ and its associate momentum. In terms of these smeared operators, the geon state reads, schematically\footnote{For a more careful treatment of the zero modes and the suppressed indices, see \cite{Louko:1998dj}. } \cite{Louko:1998dj}
\be
|\Psi_g \rangle \sim \prod_{\om_0,t_0} \exp \left(\a_{\om_0} \O_{\om_0,t_0}^\dag \O_{\om_0,-t_0}^\dag \right) |0\rangle.
\ee
Since the modes $\O_{\om_0,t_0}$ behave like free field modes at large $N$, the effect of a unitary rotation of the form \eqref{psip} or \eqref{psipp}, to leading order, is just to change the phase of the $\a_{\om_0}$ coefficients in the exponent. For simplicity, we consider an example where $\a_{\om_0} \r e^{i \varepsilon} \a_{\om_0}$  just for one particular $\om_0,t_0$. The new state $|\Psi_g' \rangle$ will then satisfy a relation of the form
\be
e^{-\b \om/2} \O^\dag_{\om} | \Psi'_g \rangle = (\O_\om + i \varepsilon \, \tilde \xi_{\om_0 t_0} (\om)\, \O_{\om_0,- t_0} ) | \Psi_g'\rangle \label{mirexc}
\ee
where we take $\varepsilon \ll 1$.

Unlike for the geon or the $J$ quotient of BTZ, we do not have a direct construction of the dual geometry for this case. However, we expect the geometry to be a small deviation from the geon. There are two possible perspectives on this modification. 

\begin{figure}[h]
\centering
\includegraphics[height=4  cm]{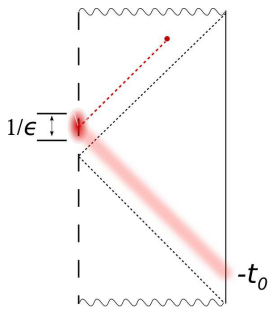} 
\caption{
If we assume $|\Psi_g' \rangle$ is an equilibrium state, the change in the geometry can be described as a modified boundary condition on the left. Rays that trace back through this point produce the indicated smeared image of the bulk mode, in agreement with the PR prediction that the locality of the mirror operator starts breaking down.}
\label{fuzzy}
\end{figure}

The first option is to apply the PR prescription treating $|\Psi_g' \rangle$ as an equilibrium state.  Translating the mirror relation \eqref{mirexc} to position space (using an computation analogous to \eqref{ftmir}), we find that for $t \neq t_0 $, the mirror of $\O(t)$ is still $\O (-t)$, but for $t \approx t_0$, the mirror is an operator smeared around $-t_0$ with width $2\pi/\e$, and thus it has become slightly non-local. Next, using the PR prescription \eqref{bulkin}, we can compute the field  $\Phi^{II}(t,r) $ behind the horizon.  By construction, the bulk field operator we obtain will perceive the late-time region as vacuum \cite{Papadodimas:2012aq}. Analytically continuing the expression \eqref{bulkin} to $t<0$,  we  find that $\Phi^{II}(t,r)$ no longer equals $\Phi^{II}(-t,r)$ for $ \pm t - r_\star \approx  t_0$, but that the relation between the two has been ``smeared''. Consequently, the reflection condition at $X_2=0$ ($t=0$) becomes ``fuzzy'' for $|r_\star| \approx t_0$, with a width of order $2\pi/\e$. This is illustrated in figure \ref{fuzzy}. Due to this ``fuzziness'', a ray that traces back through this point will be scattered to yield a boundary operator that is smeared over a range $2\pi/\e$, in agreement with the PR prediction that the mirror operator is now slightly non-local. 

The other option is to view $|\Psi_g' \rangle$ as an excitation of the geon state $|\Psi_g \rangle$. In this case, the field operator $\Phi^{II}(t,r,\phi)$ in the late-time region would be obtained from the PR prescription for the geon, i.e. it would be given by  \eqref{geonin}.  $|\Psi_g' \rangle$ itself can be interpreted as applying a unitary rotation between the field modes inside and outside the horizon. This has no effect on observations confined to one of the two regions, but it implies that $|\Psi_g' \rangle$ is not annihilated by the annihilation operators \eqref{gannih} associated with infalling observers, so they will see an excitation as they cross the horizon.

Unlike in the previous example, we do not know what the correct geometry dual to $|\Psi_g' \rangle$ is. It would be very interesting  to find a way to distinguish the two scenarios above by a computation in the dual CFT. In the case where $|\Psi'_g \rangle$ is interpreted as an equilibrium state, 
 the change in the dual geometry - just as in our previous example - is at the level of the boundary conditions on the identification surface  $X_2=0$ inside the geon black hole. It would be interesting to find out how general this picture is. Within the same framework,  it would also be interesting to explore the continuation of the expression for $\Phi$ to the entire space-time.  We  expect that the full analytic continuation is no longer possible in the modified state, which might shed some light on how the analyticity properties of the spacetime start breaking down under increasingly general unitary rotations.

\subsubsection*{Acknowledgements}

The authors are grateful to C. Aron, V. Balasubramanian, D. Harlow, A. Maloney, D. Marolf, I. Morrison, K. Papadodimas, S. Raju, A. Strominger and B. van Rees for useful discussions, and especially to M. Gaberdiel for helpful correspondence and K. Papadodimas for comments on the draft. SFR's work is supported in part by the STFC under grant ST/L000407/1. The authors  would also like to thank the Centro de Ciencias de Benasque ``Pedro Pascual'', where this work was initiated, for their hospitality and  for providing an exciting environment for discussions. This work was also supported in part by the National Science Foundation under Grant No. PHYS-1066293 and the hospitality of the Aspen Center for Physics.

\appendix

\section{Structure of the crosscap state \label{argcc}}

In this appendix, we would like to argue that  the cross-cap state satisfies \eqref{cccond}.  Mapping this relation to the plane via $z = e^{t_E+i \phi}$, it becomes 
\be
\bar{\phi}^\dag_{\bar h,h} (z,\bar z) | C \rangle = \phi_{h,\bar h} (-z,-\bar z) | C \rangle. \label{euclcond}
\ee
Let us start by working out some examples in the free boson CFT, in which the crosscap state satisfies \eqref{freecc}.
One can easily check that
\be
J(z) | C \rangle = \sum_n \frac{j_n}{z^{n+1}} |C \rangle = \sum_n \frac{(-1)^{n+1} \bar j_{-n}}{z^{n+1}} | C \rangle = \frac{1}{z^2} \sum_n \frac{\bar j_{n}}{(-z^{-1})^{n+1}} | C \rangle = \frac{1}{z^2} \bar J (-\frac{1}{z}) | C \rangle = (\bar J(-\bar z))^\dag |C \rangle.
\ee
For chiral vertex operators $\V_\a(z)= :e^{i \a X(z)}:$, we have
\be
\V_\a(z)|C\rangle = e^{\a \sum_{n>0} \frac{1}{n} j_{-n} z^n} e^{-\a \sum_{n>0} \frac{1}{n} j_{n} z^{-n}} |C\rangle =  \bar \V_\a (-\frac{1}{z})|C\rangle = (\bar{\V}_{-\a}(-\bar z))^\dag |C\rangle. \label{testver}
\ee
Let us now try to extend the validity of this formula to more general situations.
 In a rational CFT, but perhaps also more generally, primary fields in the CFT can be decomposed in terms of chiral vertex operators\footnote{This argument follows  closely the one  in \cite{Pradisi:1995pp,Bruzzo:2003tf}.  } \cite{Moore:1989vd}
\be
\phi_{h,\bar h}(z,\bar z) = \sum_{i, \bar i, f, \bar f} V_h{}^f{}_i (z) \bar{V}_{\bar h}{}^{\bar f}{}_{\bar i} (\bar z) \,  \a_{i\bar i}{}^{f \bar f} ,\label{cvod}
\ee
where the sum runs over the Virasoro primaries (assuming, for simplicity, that no extended symmetries are present). The identification $(t,\phi) \sim (-t, \phi+\pi)$ that defines the crosscap translates into $z \sim -1/\bar z$ on the plane. A priori, the theory is only defined for $|z|>1$, but one can extend it to the entire complex plane by defining
\be
T(z) = \left\{ \begin{array}{ccc} T(z) & \mbox{for} & |z| > 1, \\ && \\\frac{1}{z^4} \bar{T} \left( -\frac{1}{z}\right) &\mbox{for}& |z|<1.\end{array} \right.
\ee
One can then  use the arguments  of \cite{Cardy:1984bb} to show that $n$-point correlation functions of primary fields  in presence of a crosscap behave as $2n$-point correlation functions of purely chiral operators. Effectively, the antichiral vertex operator $\bar{V}_{\bar h}{}^{\bar f}{}_{\bar i} (\bar z)$ in \eqref{cvod} can be replaced by a chiral vertex operator $V_{\bar h}{}^{\bar f}{}_{\bar i} (-1/\bar z)$.  Introducing the spin-reversed field
\be
\bar \phi_{\bar h, h}(z,\bar z) \equiv  \sum_{i, \bar i, f, \bar f}  V_{\bar h}{}^{\bar f}{}_{\bar i}(z) {\bar V}_h{}^f{}_i (\bar z) \,  \a_{i\bar i}{}^{f \bar f}
\ee
and using the above-stated property of anti-chiral operators in front of the crosscap, we have that 
\be
\langle C | \ldots \bar \phi_{\bar h, h}^\dag(z,\bar z) |C\rangle = \langle C | \ldots \phi_{h,\bar h} (-z,-\bar z) |C \rangle.
\ee
Given that this relation holds inside any correlation function, we effectively derived the Euclidean counterpart of the Lorentzian condition \eqref{cccond}.   Similar arguments can be made when the field $\phi$ is not primary, as the chiral vertex operators can also be defined for descendants. To the extent that the decomposition \eqref{cvod} holds (at least formally) in a general CFT, then our arguments would show that \eqref{cccond}, and consequently \eqref{condo}, also hold  in the general case.

\section{Solutions of the wave equation in BTZ \label{expl}}

Let us denote the two linearly independent solutions to the wave equation \eqref{waveqn} as
\be
f^{(\pm)}_{\om,m} = \left( \frac{r}{r_+}\right)^{ 2 a}  \left(\frac{r^2}{r_+^2}-1\right)^{\pm b}  {}_2 F_1 \left(1 + a \pm b-\half \Delta, \half \Delta +  a \pm  b, 1 \pm 2 b , 1-\frac{r^2}{r_+^2} \right), \label{deffpm}
\ee
where
\be
a= \frac{i \ell m}{2 r_+}\;, \;\;\;\;\; b = \frac{i \ell^2 \om}{2 r_+} \;, \;\;\;\;\; \Delta = 1 + \sqrt{1+ \ell^2 M^2}.
\ee
Using the hypergeometric identity
\be
{}_2 F_1(a,b;c;z) = (1-z)^{c-a-b}\, {}_2 F_1 \left(c-a,c-b;\,c; z\right),
\ee
one can easily show that $(f^{(+)}_{\om,m})^\star = f^{(-)}_{\om,m}$. This set of solutions is convenient because   near the horizon $r \r r_+$, $f^{(\pm)}_{\om,m}$ have the expansion
\be
f^{(+)}_{\om,m} \sim  \, 2^{2b}\, e^{i \om r_\star} \;, \;\;\;\;\;\;\; f^{(-)}_{\om,m} \sim   \, 2^{-2b} \, e^{-i \om r_\star} ,
\ee
where $r_\star$ is the tortoise coordinate introduced in \eqref{defrst}.  The combination of the two wavefunctions \eqref{deffpm} that is normalizable at infinity is
\be
f^{(norm.)}_{\om,m}\; \propto  \; f^{(+)}_{\om,m} +   2^{4b} e^{- 2 i \d_{\om,m}}  \, f^{(-)}_{\om,m} ,
\ee
where for future convenience we have defined
\be
 e^{2 i \d_{\om,m}} \equiv  2^{4b} \,  \frac{\Gamma(-2b) \Gamma(\half \Delta-a+b) \Gamma(\half \Delta+a +b) }{\Gamma(2b) \Gamma(\half \Delta+a-b) \Gamma(\half \Delta-a -b) }.
\label{defdelta}\ee
Using the hypergeometric identities \cite{gradshteyn2007}
\bea
{}_2 F_1(a,b;c;z) &=&  \frac{\G(c) \G(b-a)}{\G(b)\G(c-a)} (-z)^{-a} {}_2 F_1(a,1-c+a;1-b+a;z^{-1}) + \non \\
&& \hspace{1cm} + \frac{\G(c) \G(a-b)}{\G(a)\G(c-b)} (-z)^{-b} {}_2 F_1(b,1-c+b;1-a+b;z^{-1}) \\
{}_2 F_1(a,b;c;z) &=& (1-z)^{-a}\, {}_2 F_1 \left(a,c-b;\,c; \frac{z}{z-1}\right)\label{idk}
\eea
we can rewrite the normalizable solution as 
\be
f_{\om,m} = c_n  \left( \frac{r}{r_+}\right)^{- 2 b-\Delta}  \left(\frac{r^2}{r_+^2}-1\right)^{b}  {}_2 F_1 \left(\half \Delta + a + b, \half \Delta -  a + b; \Delta; \frac{r_+^2}{r^2} \right), \label{normsol}
\ee
where $c_n = r_+^{-\Delta}$ is fixed by the condition that the commutatior of $\Phi$ and its conjugate momentum take the standard form. It is easy to check that $f_{\om,m}^\star = f_{\om,m}$ and, using the identity \eqref{idk}, that this wavefunction is the same as that of \cite{Papadodimas:2012aq}.

As the horizon at $r \r r_+$ is approached
\be
f_{\om,m} \r C_n \, (e^{i \om r_\star+i\d_{\om,m}} + e^{-i \om r_\star-i\d_{\om,m}}), \label{horas}
\ee
where $C_n $ is given by
\be
C_n=  c_n \, \Gamma(\Delta) \left[ \frac{\Gamma(2b) \Gamma(-2b)}{\Gamma(a-b+\half \Delta) \Gamma(-a+b+\half \Delta)\Gamma(a+b+\half \Delta) \Gamma(-a-b+\half \Delta)}\right]^\half ,\label{relcn}
\ee
and $\d_{\om,m}$ has been defined in \eqref{defdelta}. 

We can also find an expression for $f_{\om,m}(r)$ close to, but not exactly at the horizon, in the limit of large frequency. Letting $f_{\om,m}(r) = r^{-1/2} u(r)$, then it can be shown that $u(r)$ satisfies \cite{Leichenauer:2013kaa}
\be
\frac{d^2 u}{dr_\star^2} + (\om^2 - V(r)) u =0, \label{eompot}
\ee
where the effective potential $V(r)$ reads
\be
V(r) = \frac{r^2-r_+^2}{\ell^4} \left(\frac{3}{4} + M^2 \ell^2 + \frac{4 m^2 \ell^2 + r_+^2}{4r^2} \right).
\ee
For $\om >> r_+/\ell^2$, $\om >> M$ and $r< \om \ell^2$, then this potential is negligible compared to the $\om^2$ term in \eqref{eompot}, and thus the solution for $u$ is $u \propto \sim \cos (\om r_\star + \d_{\om,m})$, yielding  
\be
f_{\om,m} \propto r^{-\half}  \cos (\om r_\star + \d_{\om,m}). \label{appsolf}
\ee
The $\om$ dependence of the proportionality constant is important in computing the smeared kernel. We show in the next appendix that for a massless field, the appropriate proportionality constant is $\om^{-\frac{3}{2}}$.

\section{The smearing function \label{smearing}}

\begin{figure}[h]
\centering
\includegraphics[height=3 cm]{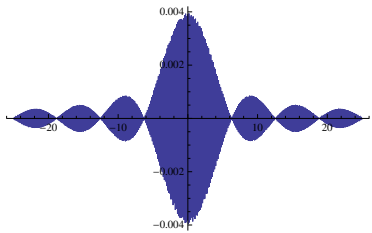}
\caption{The smeared Fourier transform of $\om^{-\frac{3}{2}} \cos \om x$ for $\om_0 = 40$, $\e =1$. } \label{locf}
\end{figure}

Here we will  evaluate the smearing kernels \eqref{smki} defined in section \ref{btzsmear}. It will  be useful to evaluate the integrals
\be
\frac{1}{\e} \int_{\om_0 - \half \e}^{\om_0 + \half \e} d \om \cos ( a \om + b ) = \cos (a \om_0 +b)  \frac{\sin (\half \e a)}{\half \e a}.
\ee
and similarly for cosine replaced by sine. As $r \r \infty$, we have $
f_{\om,m} (r) \r r^{-\Delta} $. 
Taking $\e \approx 1 $, the sum over  $m$ reduces to a single term $m_0$, and the integral yields
\be
K^I_{\om_0,t_0} (t_0-t,\e,\phi_0-\phi) = \frac{1}{2\pi^2 r^\Delta}  \cos \left[ \om_0 (t_0-t) - m_0 (\phi_0-\phi) \right] \mbox{sinc} \left(\half \e (t_0-t) \right) .
\ee
Thus, the operator is localized on the boundary around $t_0$, with width $2\pi/\e$. As we approach the horizon, the function $f_{\om,m}$ is instead given by \eqref{horas}. Taking $\Delta =2$ (massless field), we have
\be \label{fd}
f_{\om,m} \r \frac{2}{r_+^2} \left|\frac{\Gamma(2b)}{\Gamma(a+b+1) \Gamma(a-b+1)} \right|\, \cos (\om r_\star + \d_{\om,m}).
\ee
Using the fact that $a,b$ are purely imaginary, the identity $|\Gamma(i\eta)|^2 = \pi/(\eta \sinh \pi \eta)$ and  the approximation $|b| >> |a|>>1$ ($\b \om>> |m|>>1$), we find
\be
f_{\om,m} \sim \frac{2}{\ell^3\sqrt{2\pi r_+} \,\om^{3/2} } \cos (\om r_\star + \d_{\om,m}).
\ee
We also need to understand the large $\om, m$ behaviour of $\d_{\om,m}$. Using the fact that, at large $\eta \in \mathbb{R}$
\be
-i \ln \left( \frac{\Gamma(i\eta)}{\Gamma(-i \eta)} \right) = 2 \eta (\ln \eta -1) - \frac{\pi}{2} + \O(\eta^{-1}),
\ee 
we can approximate the expression \eqref{defdelta} as 
\be
\d_{\om,m} \sim - \frac{\pi}{4} + \O(m^2/\om^2) \om
\ee
yielding
\begin{equation}
K_0^I \sim \int_{\om_0-\half \e}^{\om_0 + \half \e} \frac{d\om}{\om^{3/2}} \left( \cos [\om(t_0-t+r_\star)-m_0(\phi_0-\phi)-\frac{\pi}{4}] + \cos [\om(t_0-t-r_\star)-m_0(\phi_0-\phi)+\frac{\pi}{4}]  \right)
\end{equation}
The $\om$ integral can be performed using Mathematica. The shape of the  function we obtain is plotted in figure \ref{locf}. There are two main contributions, localized at $t=t_0 \pm r_\star$. The spread on the boundary is comparable to the spread in the bulk. Adding the $r^{-\half}$ factor to $f_{\om,m}$ as in \eqref{appsolf} does not change this conclusion. 

\bigskip

\bibliographystyle{JHEP} 
\bibliography{geon}

\end{document}